\documentclass{jfm}
\usepackage{graphicx}
\usepackage{epstopdf, epsfig}
\usepackage{amsfonts,amssymb,amsmath,mathbbol}   
\usepackage{adjustbox}

\usepackage{xcolor} 
\usepackage{caption}
\usepackage{subcaption}
\usepackage{multirow}
\usepackage{multicol}
\usepackage{graphicx}
\usepackage{bm}
\usepackage{slashbox}
\usepackage{soul}
\usepackage{adjustbox}
\usepackage[normalem]{ulem}
\usepackage{placeins}

\def\We{{\it We}}
\def\Oh{{\it Oh}}

\def\Re{{\it Re}}
\def\Fr{{\it Fr}}

\def\i{{\rm i}}

\shorttitle{Splashing of non-spherical drops}
\shortauthor{N. V. Anirudh, S. Behera and K. C. Sahu}

\title{Unveiling crown-finger instability of a non-spherical drop impacting a liquid surface}

\author{Nagula Venkata Anirudh,\aff{1} Sachidananda Behera\aff{1} \corresp{\email{sbehera@mae.iith.ac.in}} \and \\ Kirti Chandra Sahu\aff{2}\corresp{\email{ksahu@che.iith.ac.in}}}

\affiliation{\aff{1} Department of Mechanical and Aerospace Engineering, Indian Institute of Technology Hyderabad, Kandi - 502 284, Sangareddy, Telangana, India
\aff{2}Department of Chemical Engineering, Indian Institute of Technology Hyderabad, Kandi - 502 284, Sangareddy, Telangana, India}

\begin{document}

\maketitle

\begin{abstract}
We present a three-dimensional numerical study of the splashing dynamics of non-spherical droplets impacting a quiescent liquid film, covering a wide range of aspect ratios ($A_r$) and Weber numbers ($\We$). The simulations reveal distinct impact dynamics, such as spreading, splashing type-1, splashing type-2, and canopy formation, which are delineated in a regime map constructed in the $A_r$–$\We$ parameter space. Our results demonstrate that droplet morphology during the impact significantly influences crown evolution and splash initiation, with oblate drops promoting finger growth and fragmentation due to enhanced rim deceleration, while prolate drops tend to form canopies. We observe that the hole instability, which becomes more prominent at higher Weber numbers, arises from lamella rupture in the thinnest region of the film, located just beneath the crown rim. A linear stability analysis, supplemented by the temporal evolution of the crown obtained from the numerical simulations, adequately predicts the number of fingers formed along the crown rim by accounting for both Rayleigh–Plateau (RP) and Rayleigh–Taylor (RT) instabilities. The theoretical analysis demonstrates the dominant role of the Rayleigh-Plateau instability in determining the number and wavelength of early undulations, with the Rayleigh-Taylor instability serving to amplify the growth rate of the disturbances. Our findings highlight the critical role of the droplet shape in splash dynamics, which is relevant to a range of applications involving droplet impact.
\end{abstract}



\section{Introduction} \label{sec:intro}

The study of impacts of drops on surfaces has a long and distinguished history, beginning with Leonardo da Vinci’s observations in the Codex Leicester (1500). The seminal work of Worthington \citep{worthington1877xxviii, worthington1908study} on the complex dynamics of droplet impact on a liquid interface laid the foundation for modern research in this field. Droplet impact on a surface is central to a wide range of industrial and environmental processes, including spray cooling, inkjet printing, fuel injection, healthcare, and agriculture, as well as microfluidics, combustion, coatings, soil erosion, air entrapment at the sea surface, and water–oil separation in petroleum recovery, to name a few \citep{stone2004engineering, thoroddsen2008high, kavehpour2015coalescence}. Furthermore, understanding the impacts of droplets is vital for interpreting natural phenomena such as raindrop collisions with liquid surfaces and the ejection of marine microplastics from lakes and oceans \citep{low1982collision, thomson1886v, pumphrey1989underwater, veron2015ocean, liu2018experimental}.

Droplet impact on a liquid surface triggers several distinct phenomena, such as crater formation, air bubble entrainment, jetting, and splashing \citep{thoraval2012karman, wang2013we, deka2017regime, behera2023investigation}. These processes are governed by the interplay of inertial, surface tension, and viscous forces and are influenced by parameters such as impact velocity, liquid film thickness, and fluid properties. The Weber number ($\We = {\rho_l U_0^2 R_{eq}/\sigma}$), which represents the ratio of inertial forces to surface tension, and the Reynolds number ($\Re = {U_0 \rho_l R_{eq}/\mu_l}$), representing the ratio of inertial forces to viscous forces, are commonly used to characterize the transition between spreading and splashing, as well as the morphology of the resulting crown structures \citep{gao2015impact, roisman2006spray}. Here, $U_0$ is the impact velocity, $R_{eq}$ is the equivalent radius of the drop, $\sigma$ is the interfacial tension, and $\rho_l$ and $\mu_l$ are the density and dynamic viscosity of the liquid, respectively. In addition, the Ohnesorge number ($\Oh = \mu_l / \sqrt{\rho_l \sigma R_{eq}} = \sqrt{\We}/\Re$) and the Froude number ($\Fr = U_0^2 / g R_{eq}$) are employed to characterize different regimes of droplet impact dynamics \citep{dandekar2025splash, murphy2015splash}. The Ohnesorge number quantifies the relative importance of viscous forces compared to the combined effects of inertia and surface tension, while the Froude number represents the ratio of inertial to gravitational forces.

Among the various phenomena observed during droplet impact on a liquid surface, splashing, where a thin radial liquid sheet breaks up into secondary droplets, is one of the most extensively studied yet intricate processes, owing to its crucial role in fluid dispersion and energy dissipation \citep{cossali1997impact, wang2000splashing, thoroddsen2002ejecta, josserand2003droplet, gao2015impact, wang2023analysis, dandekar2025splash}. Splashing unfolds in distinct stages, each characterized by morphological changes that emerge after the initial impact \citep{weiss1999single, thoroddsen2002ejecta, agbaglah2015drop, josserand2003droplet, ray2012oblique, ray2015regimes, gielen2017oblique, okawa2008effect}. For a spherical drop impacting a liquid surface, the process begins with the droplet spreading radially, forming a thin lamella. As the lamella expands, a raised rim forms at the edge driven by inertia, while surface tension tends to stabilize the structure. The rim then undergoes Rayleigh-Plateau instabilities, creating nodes along the edge of the crown, which indicates the transition to finger formation \citep{roisman2006spray, yarin1995impact}. Subsequently, the fingers elongate and break into secondary droplets due to Rayleigh-Taylor (RT) and Rayleigh-Plateau (RP) instabilities \citep{josserand2016drop, constante2023impact}. The residual liquid film either retracts or spreads further until equilibrium is reached, influenced by surface tension and viscous damping. Higher Weber numbers can trigger prompt splashing \citep{wang2023analysis, thoroddsen2002ejecta}. Splashing is often modelled as a kinematic discontinuity \citep{yarin1995impact}, where a velocity jump in the liquid film propagates outward, controlling the number of fingers and secondary droplets. 

To investigate the temporal evolution of crown rim dynamics resulting from the impact of a spherical drop on a liquid surface, \citet{roisman2006spray}, \citet{agbaglah2013longitudinal}, and \citet{agbaglah2014growth} conducted linear stability analyses. Their studies showed that the most unstable mode predicted by the theory can be used to estimate both the number and timescale of secondary droplet formation during splashing. The splashing dynamics also depend on whether the impact occurs with a thin liquid layer or a deep pool. A Worthington jet is generated upon impact with a thin liquid layer \citep{gekle2010, gordillo2010, josserand2003droplet}, while a ``milkdrop coronet'' forms in deep liquid pools \citep{josserand2016drop, constante2023impact}. At high Weber numbers, capillary waves are absent; instead, a lamella is ejected from the droplet front \citep{leneweit2005regimes, liu2016numerical}. The curvature in the contact line region, influenced by surface tension or stagnation pressure, dictates whether capillary waves form or a lamella emerges. Previous theoretical studies have examined interfacial instabilities in accelerating and curved geometries in considerable depth. \citet{krechetnikov2009crown} analyzed the stability of curved interfaces and showed that curvature can either suppress or amplify Rayleigh–Taylor growth, depending on its orientation. Building on this, \citet{krechetnikov2010stability} investigated the coupled Rayleigh–Taylor and Rayleigh–Plateau mechanisms in accelerating liquid sheets, demonstrating the critical role of unsteady acceleration in rim destabilization. Subsequently, \citet{krechetnikov2017stability} and \citet{krechetnikov2017stabilityblob} extended this framework to time-dependent and globally evolving base states, emphasizing that when the interface evolution occurs on a timescale comparable to the instability growth, the classical quasi-steady linear theory is no longer valid. Most recently, \citet{krechetnikov2024transverse} examined nonlinear modal interactions and wave dynamics on concentric interfaces, highlighting important post-linear effects. Collectively, these contributions offer a comprehensive theoretical foundation for understanding the early evolution of crown rims, where curvature, rim deceleration, and liquid-layer thinning jointly dictate the fingering patterns that emerge during drop impact.

A few more interesting phenomena, namely hole instability and canopy formation during the rising lamella, are also observed under certain conditions. The hole instability, characterized by the formation of a cavity within the crown upon impact with a liquid surface, arises from the displacement of liquid at the impact point, resulting in a depression while the surrounding fluid is pushed outward to form the crown rim. For spherical drops, studies by \cite{josserand2003droplet} and \cite{thoroddsen2002ejecta} emphasize the critical role of hole dynamics in determining the size and number of secondary droplets. Recent experimental and theoretical studies by \cite{dandekar2025splash} revealed fascinating dynamics, including canopy splashing during impacts on confined liquid surfaces \citep{deka2017regime} and deep pools \citep{dandekar2025splash, worthington1908study}. \cite{dandekar2025splash} also demonstrated that the interplay between interfacial and gravitational forces governs the formation of crown splashes. They further showed that crater dynamics depend on impact speed and droplet size. A similar observation was also previously reported by \cite{bisighini2010crater}.

It is worth noting that most previous studies have focused on spherical droplets, while only a few investigations \citep{deka2017regime, anirudh2024coalescence, khan2024impact} have explored the coalescence dynamics of non-spherical droplets on liquid surfaces. For instance, \citet{khan2024impact} conducted experiments on the impact of non-spherical droplets on a thin liquid film and found that oblate droplets are more likely to undergo splash-induced fragmentation during impact. \cite{deka2017regime} performed axisymmetric numerical simulations to investigate large bubble entrapment in prolate-shaped drops, attributing this phenomenon to their deeper penetration of the resulting vortex ring into the liquid pool. \cite{anirudh2024coalescence} conducted three-dimensional simulations to study the transition between partial and complete coalescence in non-spherical drops placed on a liquid surface. Their findings revealed that non-spherical drops with aspect ratios, $A_r$ (the ratio between the horizontal and vertical axes) greater than 0.67 experience partial coalescence, forming a secondary/daughter droplet. They also observed that oblate drops with high aspect ratios trap air in a ring-like bubble, which forms a liquid column that undergoes Rayleigh-Plateau instability, resulting in two daughter droplets. However, these studies primarily focus on coalescence and air entrapment rather than splashing or crown dynamics. Furthermore, it is interesting to note that the impact of non-spherical droplets can significantly modify the development of hole instability and influence canopy splashing, owing to distinct rim and crown dynamics governed by their shape at the moment of impact. Non-spherical drops also introduce additional complexities, such as anisotropic dynamics, asymmetric rim formation, and shape-dependent instability growth. Despite these factors, the influence of droplet shape on splashing dynamics remains largely unexplored, warranting further investigation.

Thus, the present study aims to bridge this gap by investigating the splashing dynamics of non-spherical drops impacting liquid surfaces through three-dimensional numerical simulations. Various collision outcomes, such as spreading, splashing type-1, and splashing type-2, are examined for non-spherical drops and compared with those observed in spherical drops. Additionally, we investigate the dynamics of canopy formation and the influence of droplet shape on hole instabilities, which lead to the development of distinct fingering patterns and subsequently detach into secondary droplets at different Weber numbers. A comprehensive regime map is constructed by performing a large number of simulations across a range of aspect ratios ($A_r$) and Weber numbers ($\We$), demarcating the transition from spreading to splashing type-1 to splashing type-2 behaviours. We observe that the hole instability, which becomes more pronounced at higher Weber numbers, originates from lamella rupture in the thinnest region of the film just beneath the crown rim. Furthermore, our study presents a novel theoretical framework to predict the number of crown fingers based on the most unstable wavelength obtained from linear stability analysis, extending beyond existing analytical models developed for spherical droplets \citep{roisman2006spray, agbaglah2013longitudinal, agbaglah2014growth}. 

The remainder of the manuscript is structured as follows. Section~\S\ref{sec:method} describes the problem formulation, outlines the numerical methodology used for the three-dimensional simulations, and presents the validation of the flow solver. Section~\S\ref{sec:dis} discusses the simulation results and introduces a linear stability analysis to investigate the temporal evolution of crown rim dynamics and predict the number of fingers based on the most unstable mode. Finally, Section~\S\ref{sec:conc} provides the concluding remarks.

\section{Formulation} \label{sec:method}

\begin{figure}
\centering
\includegraphics[scale=0.5]{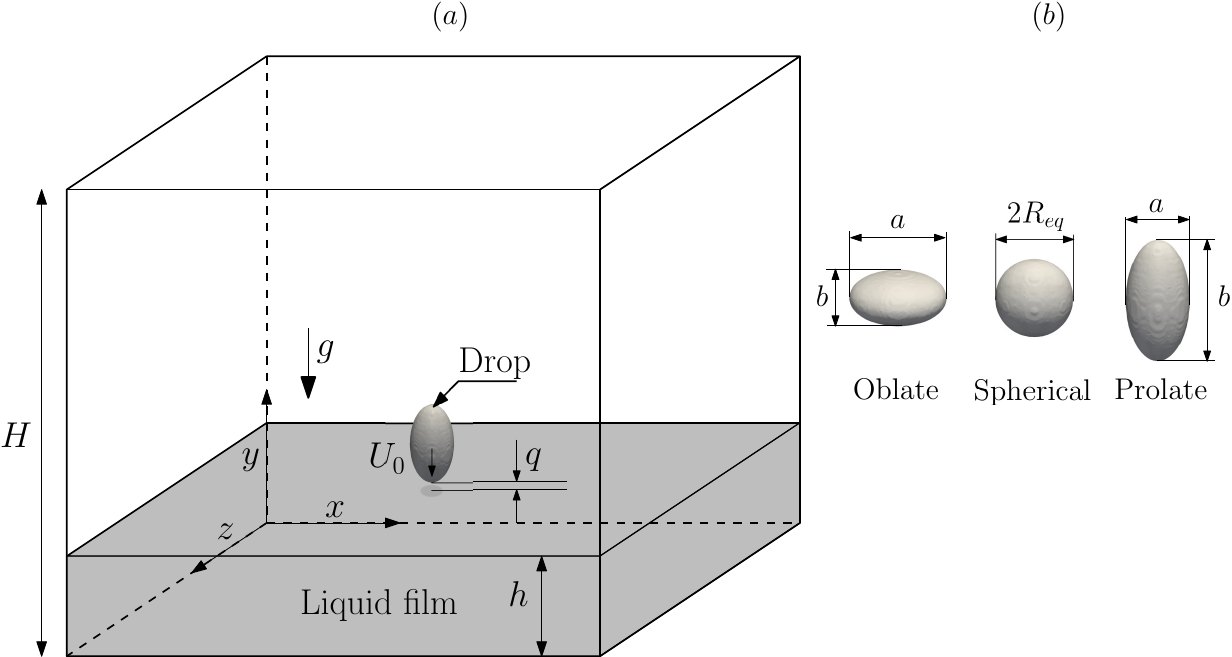}
\caption{(a) Schematic representation of the three-dimensional computational domain ($H \times H \times H$) illustrating the initial configuration of a non-spherical drop impacting a liquid pool with velocity $U_0$. (b) Illustration of the various initial shapes of the non-spherical drop, classified as oblate ($A_r = a/b > 1$), spherical ($A_r = 1$), and prolate ($A_r < 1$). Here, $R_{eq}$ denotes the equivalent spherical radius of the drop, $h$ represents the depth of the liquid film, and $q$ is the separation distance between the bottom of the drop and the liquid film.}
\label{fig:schematic}
\end{figure}

The dynamics of a non-spherical drop impacting a liquid surface is examined through three-dimensional numerical simulations of the Navier-Stokes and continuity equations. The initial configuration is schematically illustrated in figure \ref{fig:schematic}(a), while figure \ref{fig:schematic}(b) depicts the drop shapes at the moment of impact. The computational domain is a three-dimensional cube with dimensions $H \times H \times H$, where $H = 20 R_{eq}$. Here, $R_{eq} = (a^2 b / 8)^{1/3}$ represents the equivalent spherical radius of non-spherical droplets, defined to ensure a constant droplet volume. A Cartesian coordinate system $(x, y, z)$ is employed, with gravity $(g)$ acting in the negative $y$ direction. At time $t = 0$, the drop is placed at $q = 0.09 R_{eq}$ from the free surface and given an initial velocity $U_0$ in the negative direction $y$, while the rest of the region in the computational domain remains stationary. The height of the liquid pool is set to $h = 2 R_{eq}$ in the present study. The diameters of the non-spherical drop along the horizontal $(x)$ and vertical $(y)$ axes are denoted by $a$ and $b$, respectively. We assume that the diameter of the drop in the spanwise $(z)$ direction is the same as that in the horizontal $(x)$ direction. A large computational domain is used to ensure that boundary effects are negligible during the simulations. We define $A_r (=a/b) > 1$, $A_r = 1$, and $A_r < 1$ as oblate, spherical, and prolate drops, respectively (see, figure \ref{fig:schematic}b). Both the drop and the pool consist of the same liquid, with air as the surrounding medium. All fluids are considered to be Newtonian and incompressible, with the dynamic viscosities and densities of the liquid and air denoted by $(\mu_l, \rho_l)$ and $(\mu_a, \rho_a)$, respectively. 

\subsection{Dimensionless governing equations}

We employ the following scaling to render the governing equations dimensionless: 
\begin{eqnarray} (x,y,z) = R_{eq} (\widetilde x,\widetilde y,\widetilde z), ~ t =  (R_{eq}/U) \widetilde t, ~ \bm{u} = U \bm{\widetilde u}, ~ P = (\rho U^2) \widetilde P, \nonumber \\ \mu = {\widetilde \mu} \mu_l, ~ \rho = {\widetilde \rho} \rho_l, ~ {\rm and} ~ \delta_s = {\widetilde \delta_s} R_{eq}. ~~~~~~~~~~~~~~\label{scale} 
\end{eqnarray} 
Here, the tildes designate dimensionless quantities; $\bm{u} = (u,v,w)$ is the velocity field, where $u$, $v$, and $w$ denote the components of $\bm{u}$ in the $x$, $y$, and $z$ directions, respectively; $P$ represents the pressure field; and $\delta_{s}(\bm{x}-\bm{x}_f)$ is the delta distribution function, which is zero everywhere except on the interface ($\bm{x}=\bm{x}_f$), where $\delta_s=1$. After droppling the tilde notations from all non-dimensional variables, the dimensionless continuity and Navier–Stokes equations governing the impact dynamics of the drop onto the liquid pool are expressed as
\begin{subequations}
\begin{eqnarray}
\nabla \cdot \bm{u} &=& 0, \label{eqn:eq1} \\
\rho \left(\frac{\partial \bm{u}}{\partial t} + \bm{u}. \nabla \bm{u} \right) &=& -\nabla P + \frac{1}{Re} \nabla \cdot \left [\mu (\nabla \bm{u} + \nabla \bm{u}^T) \right]  + \frac{1}{\We}\kappa \bm{n} \delta_s  - \frac{\rho g}{\Fr}  \hat{j},  ~~~~  ~~\label{eqn:eq2}
\end{eqnarray}
\end{subequations}
where $\hat{j}$ represents the unit vector in the vertical (negative $y$) direction, and $\kappa~(\equiv - \nabla \cdot n)$ denotes the interfacial curvature, with $n$ being the outward-pointing unit normal to the interface. It is important to note that the surface tension force is included as a body force term in eq. (\ref{eqn:eq2}) using the continuum surface force (CSF) formulation proposed by \cite{brackbill1992continuum}.

The dynamic interface between the air and liquid phases is captured by solving an advection equation for the volume fraction of the liquid phase, denoted by $c$, within the framework of the Volume of Fluid (VOF) approach. For the air and liquid phases, $c = 0$ and $c = 1$, respectively, and the advection equation for $c$ is given by:
\begin{eqnarray} 
{\frac{\partial c}{\partial t}} + \bm{u} \cdot \nabla c = 0. 
\label{eqn:VOF} 
\end{eqnarray}
The density field $(\rho)$ and the viscosity field $(\mu)$ can be expressed as:
\begin{subequations}
\begin{eqnarray} 
\rho = \rho_r (1-c) + c , 
\label{eq_rho} \\
\mu = \mu_r (1-c) + c , \label{eq_mu} 
\end{eqnarray}
\end{subequations}
where $\rho_r (=\rho_a/\rho_l)$ and $\mu_r (=\mu_a/\mu_l)$ represent the density and viscosity ratios, respectively. Unless stated otherwise, pure water is used as the working liquid, with air as the surrounding medium. The physical properties of water are $\mu_l = 0.001$~Pa$\cdot$s, $\rho_l = 1000$~kg/m$^3$, and $\sigma = 0.072$~N/m. Consequently, for the air-water system, the viscosity and density ratios are $\mu_r = 1.48 \times 10^{-2}$ and $\rho_r = 10^{-3}$, respectively. In the present study, we simulate the impact of a non-spherical droplet with an equivalent radius of $R_{eq} = 2.1$~mm on a quiescent liquid surface. The impact velocity of the droplet is varied to change the relevant dimensionless numbers, such as the Weber number $(\We)$, Reynolds number $(\Re)$, and Froude number $(\Fr)$.

\subsection{Numerical method}

We employ a two-phase flow solver within the OpenFOAM framework to simulate the impact dynamics of a non-spherical droplet on a liquid pool. The governing equations are implemented on a collocated grid, where velocity components, pressure, and volume fraction values are defined at the cell centers. To track the interface between the liquid and gas phases, the volume of fluid (VOF) method is used, with a sharp interface maintained for accurate surface tension force calculations as described in eq. (\ref{eqn:eq2}) \citep{brackbill1992continuum}. To achieve this, a uniform orthogonal mesh is employed along with an adaptive mesh refinement (AMR) strategy to ensure precise capture of interfacial dynamics while avoiding excessive computational cost. The AMR is guided by the spatial variation in the volume fraction field ($c$), with mesh refinement applied in interfacial cells where $0 < c < 1$ to accurately resolve crucial dynamic features, such as the expanding crown, secondary droplet ejection, and thin liquid sheet formation. Using this mesh refinement strategy, the cells are coarsened away from the interface, thereby enhancing computational efficiency. The mesh refinement level is set to three, resulting in a mesh size ranging from 7 to 14 million cells. 

\begin{figure}
\centering
 \includegraphics[width=0.4\linewidth]{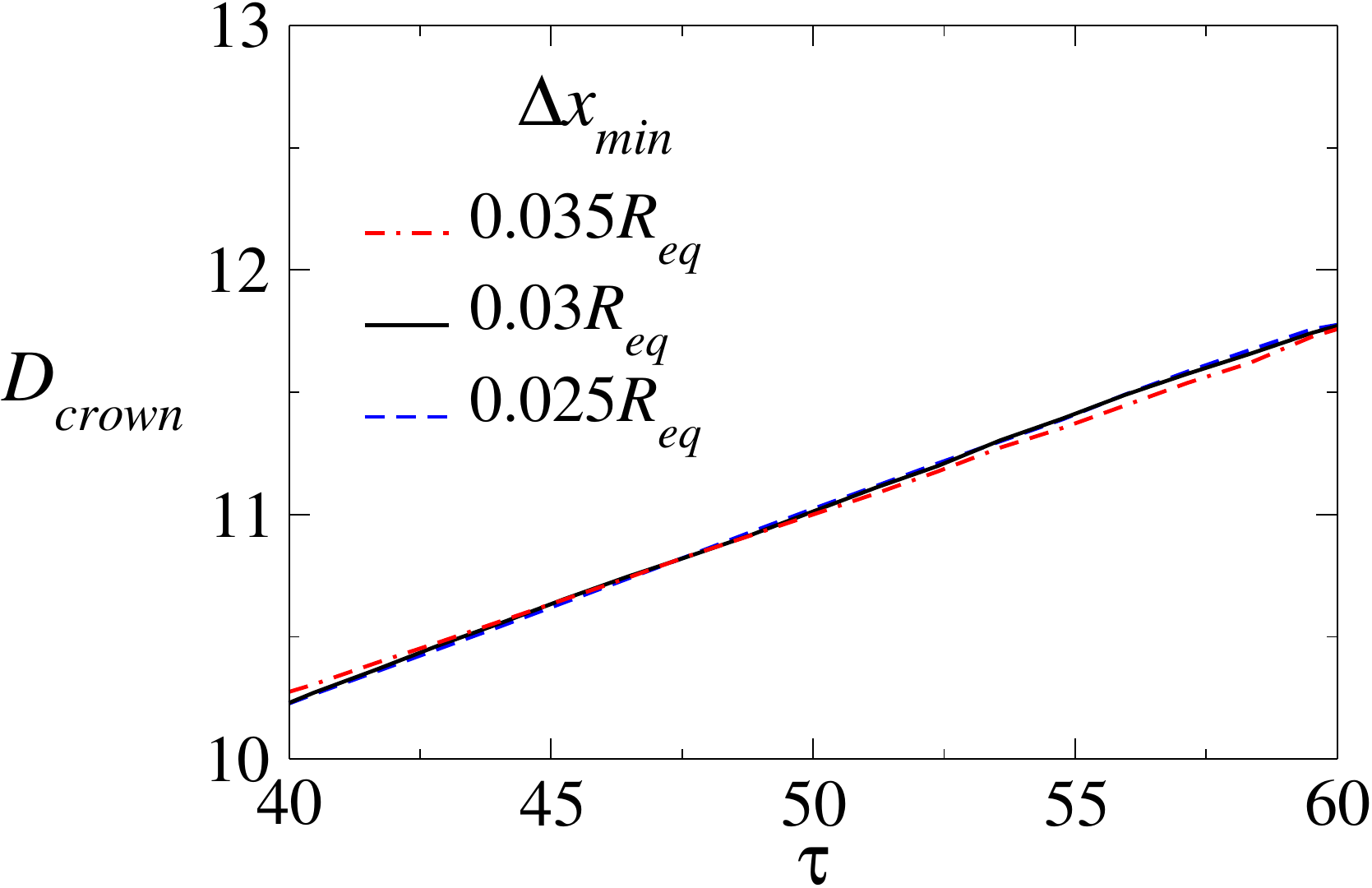}
 \caption{Grid-independence test showing the temporal evolution of the crown diameter $D_{{crown}}$ for three minimum grid sizes at $\We = 729$ and $A_r = 1.5$.}
  \label{fig:grid_test}
\end{figure}

\begin{figure}
\centering
\includegraphics[scale=0.65]{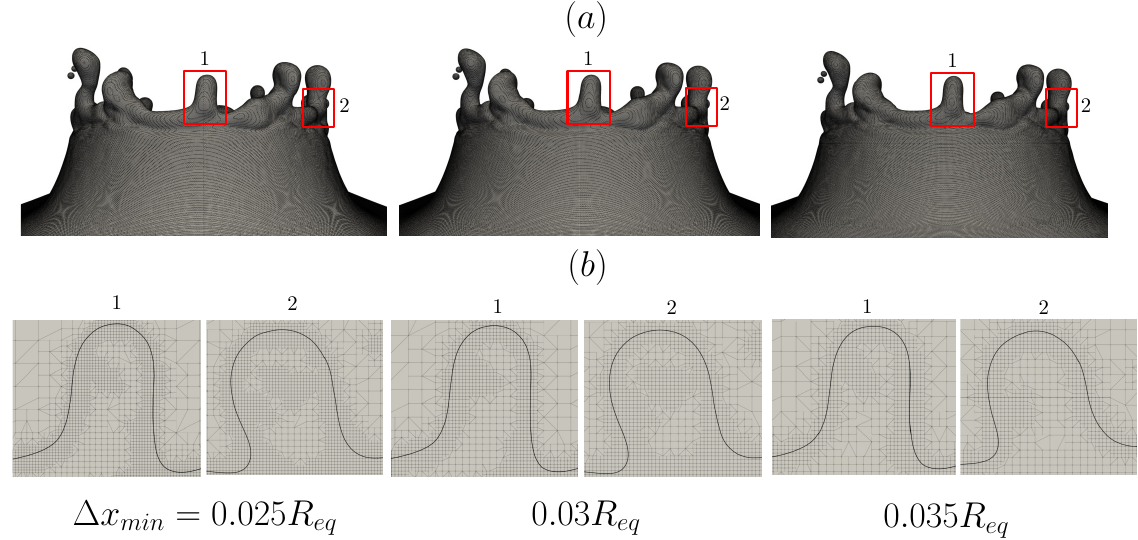}
\caption{Grid-independence test: Panel ($a$) shows the crown morphology for three minimum grid sizes, $\Delta x_{\min} = 0.035R_{eq}$, $0.03R_{eq}$, and $0.025R_{eq}$, at $\We = 729$ and $A_r = 1.5$ at $\tau = 42.86$. Panel ($b$) presents a magnified view of a representative finger for all three grids.}
\label{fig:grid_indep_mp}
\end{figure}

\begin{figure}
\centering
\includegraphics[width=1\linewidth]{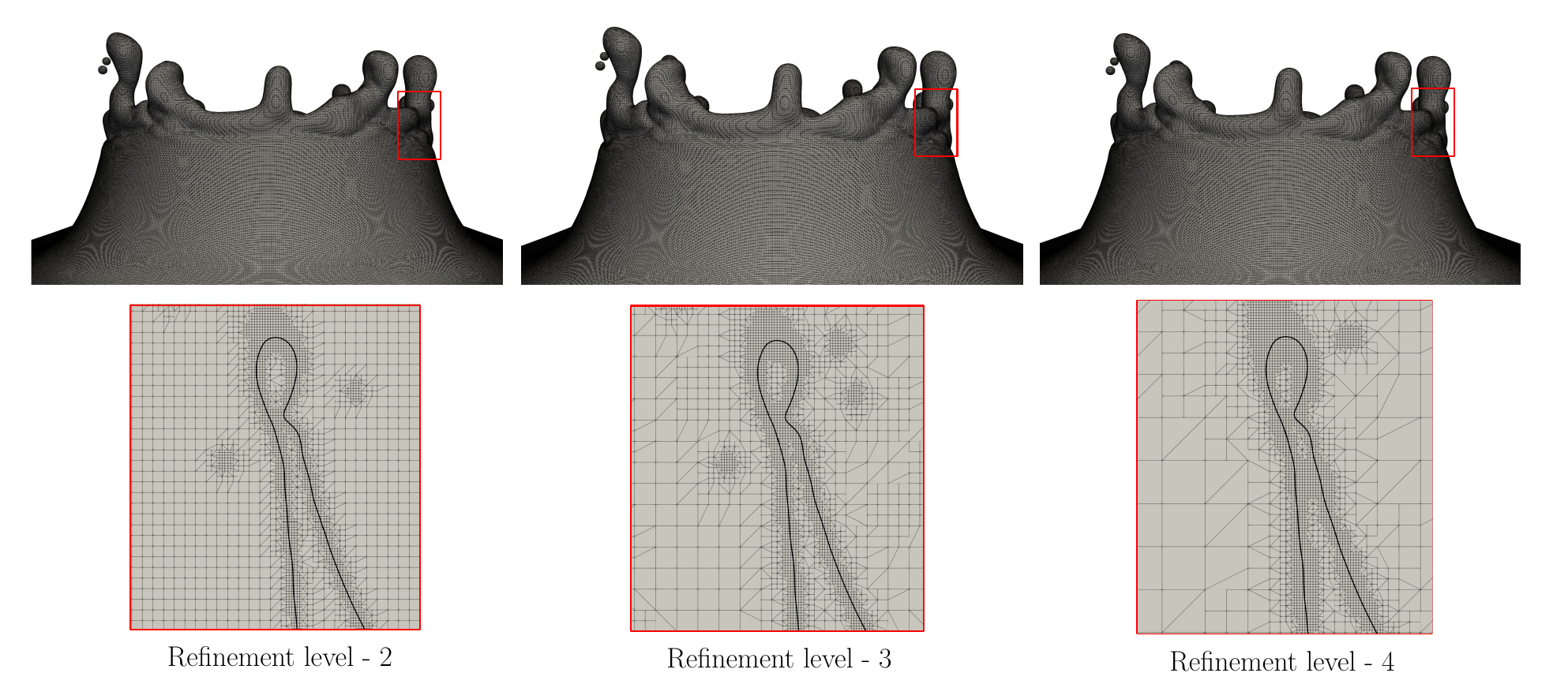}
\caption{Effect of the adaptive mesh refinement (AMR) parameter on the crown morphology for $\We = 729$ and $A_r = 1.5$ at $\tau = 42.86$. The bottom panels display the cross-sectional front view of the crown in the $x$–$z$ plane along with the corresponding computational grids for each refinement level.}
\label{fig:amr_test}
\end{figure}

\begin{figure}
\centering
\includegraphics[width=0.4\linewidth]{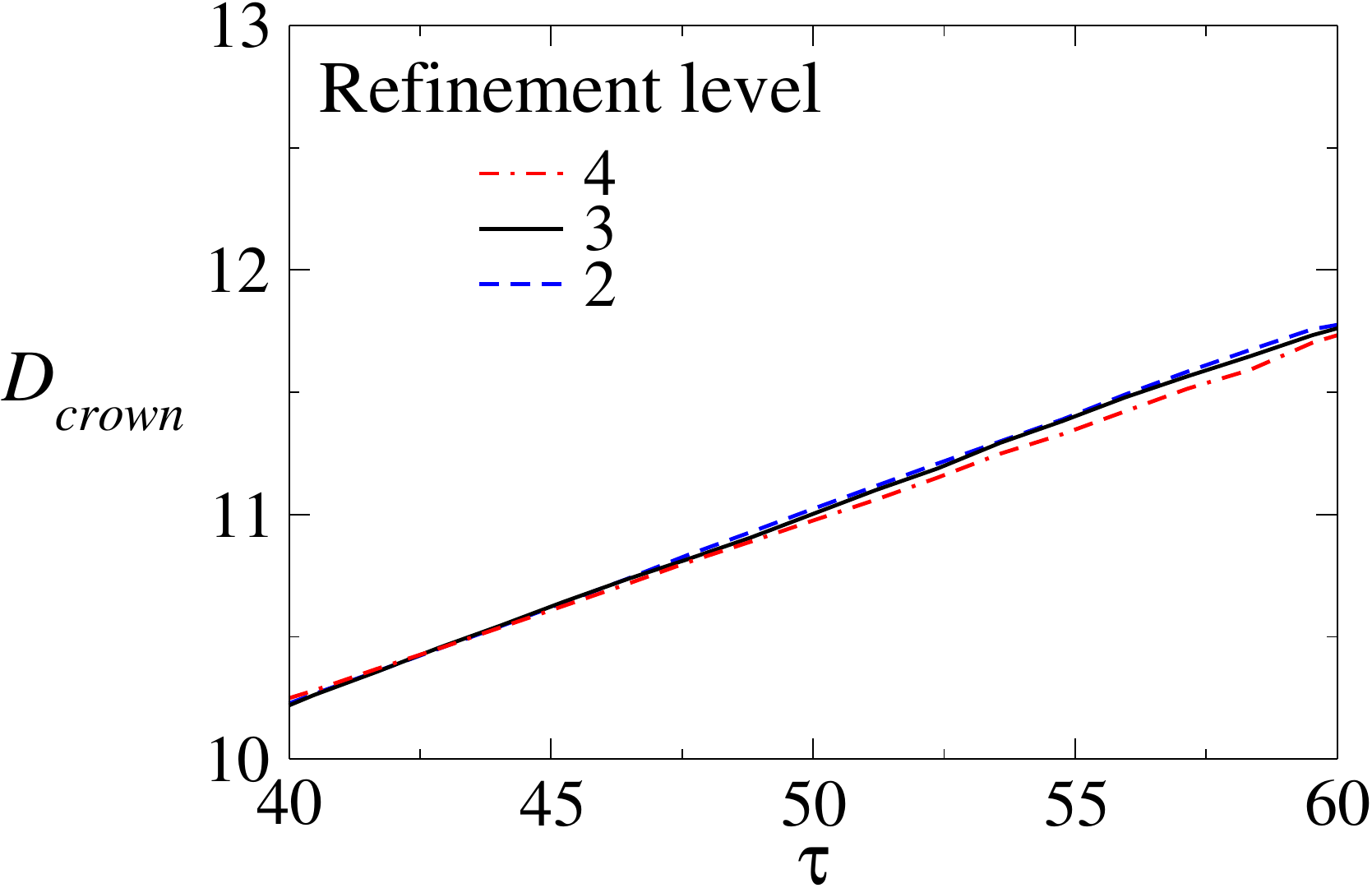}
\caption{Effect of the adaptive mesh refinement (AMR) parameter on the temporal evolution of the crown diameter $D_{{crown}}$ for $\We = 729$ and $A_r = 1.5$.}
\label{fig:amr_quant_test}
\end{figure}

Figure \ref{fig:grid_test} presents the temporal evolution of the crown diameter $D_{{crown}}$ for three grid sizes ($\Delta x_{\min} = 0.035R_{eq}$, $0.03R_{eq}$, and $0.025R_{eq}$) at $\We = 729$ and $A_r = 1.5$. It can be seen that the temporal variation of $D_{crown}$ for the three grids overlaps closely, with the deviation between the two finest meshes ($\Delta x_{\min} = 0.03R_{eq}$ and $0.025R_{eq}$) remaining below $1\%$ throughout the simulation. Additionally, the crown and finger structures shown in figure \ref{fig:grid_indep_mp} depicts negligible difference across the grids, confirming that both the global crown dynamics and rim-scale instabilities are fully converged at $\Delta x_{\min} = 0.03R_{eq}$. To examine the sensitivity to the adaptive mesh refinement (AMR) parameters, we performed additional tests using three refinement levels of 2, 3, and 4 for the same representative case ($A_r = 1.5$, $\We = 729$). At $\tau = 42.86$, the crown morphology obtained from all three refinement levels is visually identical (figure \ref{fig:amr_test}), and the temporal evolution of $D_{crown}$ shows almost perfect overlap (figure ~\ref{fig:amr_quant_test}). The deviation in $D_{crown}$ between the two finest AMR levels remains below $1\%$ throughout the evolution, demonstrating that the results are insensitive to further refinement. Based on these results, we adopt $\Delta x_{\min} = 0.03R_{eq}$ and AMR level 3 for all simulations as an optimal balance between accuracy and computational cost.

Furthermore, to mitigate interface smearing in the VOF method, we used the numerical interface compression technique using the multidimensional universal limiter with explicit solution (MULES). This approach introduces an additional term to the governing equation, sharpening the interface between the two fluids and improving the accuracy of the simulation. The modified governing equation for the volume fraction $c$ is given by:
\begin{equation} 
\frac{\partial c}{\partial t} + \nabla \cdot (\bm{u} c) - \nabla \cdot (\bm{u}_{r,f} c(1-c)) = 0, 
\end{equation}
where the term $\bm{u}_{r,f} c(1 - c)$ is responsible for compressing the interface, ensuring that $c$ remains bounded between 0 and 1 within a thin interfacial region. The relative velocity between the two fluids ($\bm{u}_{r,f}$) can be expressed as:
\begin{equation} 
\bm{u}{r,f} = \min \left( C{\text{compr}}, \frac{|\phi_f|}{|S_f|}, \max \left[\frac{|\phi_f|}{|S_f|} \right] \right) \bm{n}_f, 
\end{equation}
where $C_{\text{compr}}$ is the user-defined interface compression factor, which is set to 1 in our study, as recommended by \cite{deshpande2012evaluating}. This setting helps prevent excessive interface diffusion while ensuring numerical stability. The term $\phi_f$ represents the volumetric flux across a computational cell face, $S_f$ denotes the outward-pointing area vector at that face, and $\bm{n}_f$ is the unit normal vector at the face center, pointing along the local interface normal. By incorporating this interface compression approach, the numerical method maintains a well-defined liquid-gas boundary, preventing artificial diffusion that could otherwise affect the accuracy of the impact dynamics. For a more detailed discussion on the numerical implementation of the MULES method, the reader is referred to \cite{deshpande2012evaluating}. The following boundary conditions are employed in our simulations: a free-slip condition is applied on all side boundaries, while the bottom wall is subjected to no-slip and no-penetration conditions. In our simulations, the bottom of the pool is treated as a solid wall with a no-slip velocity boundary condition. For pressure, zero-gradient boundary conditions are applied on the side and bottom walls, while the top boundary is set to atmospheric pressure. For the volume fraction, zero-gradient conditions are used on the side and bottom walls, and the top boundary is defined as an inlet–outlet with an inlet value of zero. In the next section, we validate the solver by comparing the results obtained from the present numerical method with the experimental observations of \cite{dandekar2025splash} and \cite{wang2000splashing} for the impact of a spherical droplet on a liquid surface.

\subsection{Validation}

\begin{figure}
\centering
\includegraphics[scale=0.6]{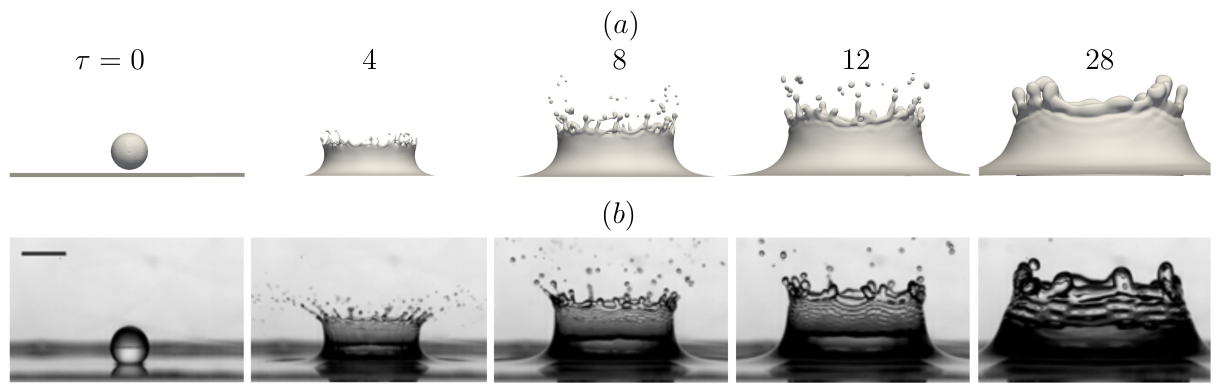}
\caption{Comparison of the impact dynamics of a spherical water droplet (initial diameter $d_0 = 4.2$ mm) on a deep water pool obtained from (a) the present numerical simulations and (b) the experimental observations of \cite{dandekar2025splash}. The corresponding dimensionless parameters based on the current formulation are $\We = 495.1$, $\Re = 8652$, and $\Fr = 823.96$. The scale bar shown in panel (b) denotes a length of 5 mm.} 
\label{fig:dandekar_Validation}
\end{figure}

Figure \ref{fig:dandekar_Validation} illustrates the temporal evolution of the collision dynamics of a spherical water droplet (initial diameter, $d_0 = 4.2$ mm) impacting a deep water pool, resulting in the formation of a cylindrical splash crown. Figure \ref{fig:dandekar_Validation}(a) and \ref{fig:dandekar_Validation}(b) compare the results from the present numerical simulation and the experimental observations of \cite{dandekar2025splash}, respectively. Based on the current formulation, the associated dimensionless parameters are $\We = 495$, $\Re = 8652$ and $\Fr = 824$. Inspection of figure \ref{fig:dandekar_Validation} reveals that the present numerical simulation accurately captures the temporal evolution of crown formation, closely matching the experimental snapshots at various normalized time instances ($\tau = t U_0 / R_{eq}$), where $\tau = 0$ represents the moment the droplet first contacts the free surface. As splashing progresses, a cylindrical crown forms, shedding droplets along its rim, with most secondary droplets ejected before the crown reaches its maximum height. Thus, it can be concluded that the present numerical solver is capable of capturing complex interfacial dynamics, including crown propagation, droplet detachment, and the formation of fingers at later stages.

\begin{figure}
\centering
\includegraphics[scale=0.6]{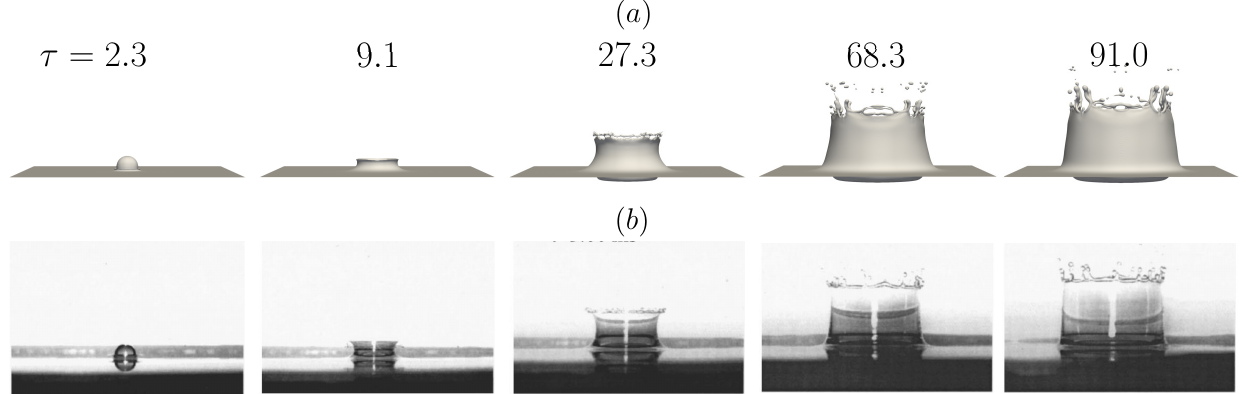}
\caption{Comparison of the impact dynamics of a spherical droplet (initial diameter $d_0 = 4.2$ mm) of 70\% glycerol-water solution on a deep pool containing the same liquid obtained from (a) the present numerical simulations and (b) the experimental observations of \cite{wang2000splashing}. The corresponding dimensionless parameters based on the current formulation are $\We = 1005$, $\Re = 585$ and $\Fr = 1262$, $\rho_r = 8.3 \times 10^{-4}$, and $\mu_r = 6.7 \times 10^{-4}$.}
\label{fig:wang_chen}
\end{figure}

\begin{figure}
\centering
\includegraphics[scale=0.3]{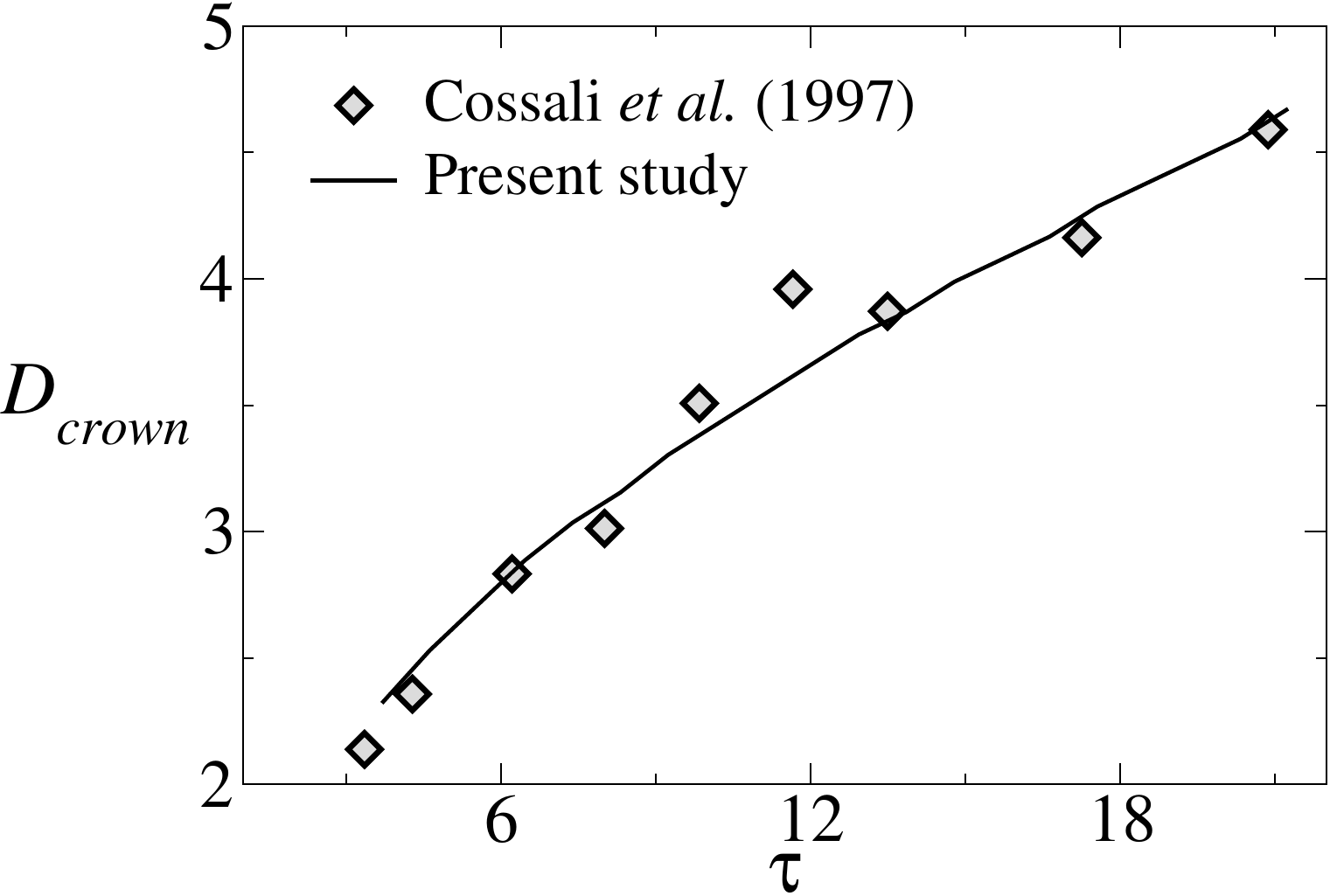}
\caption{Comparison of the temporal evolution of the normalized crown diameter, $D_{crown}$ (measured at the crown tip and normalized by the equivalent radius $R_{eq}$), obtained from the present numerical simulations with the experimental results of \citet{cossali1997impact}. The fluid system corresponds to a water–air interface, and the associated dimensionless parameters based on the current formulation are $\We = 160.1$, $\Re = 4159$, and $\Fr = 522.4$.}
\label{fig:cossali}
\end{figure}

In figure \ref{fig:wang_chen}, we compare the results obtained from the present numerical simulation with the experimental observations of \cite{wang2000splashing}, who investigated the impact of a droplet of 70\% glycerol–water solution ($d_0 = 4.2$ mm) on a deep pool of the same liquid. The comparison reveals that, for the given parameters, both the simulation and the experiments show the crown wall developing nearly perpendicular to the free surface, forming a structure resembling a circular cylinder. The temporal evolution of the crown structure observed in the simulation closely matches the experimental results, demonstrating the accuracy of the present numerical approach. Notably, the simulation effectively captures key features of the splashing dynamics, including crown expansion, secondary droplet formation, and rim instabilities. In addition, figure \ref{fig:cossali} presents a quantitative comparison of the normalized crown diameter ($D_{crown}$) over time with experimental data from \cite{cossali1997impact}, further confirming the reliability of the present simulation. For additional validation of our solver, the reader is referred to our previous study \citep{anirudh2024coalescence}.

\section{Results and discussion} \label{sec:dis}

The primary aim of this study is to investigate the splashing dynamics of non-spherical drops impacting a stationary liquid film, emphasizing the influence of drop aspect ratio ($A_r$) and Weber number ($\We$) on splash morphology. We identify four distinct impact phenomena, spreading, splashing type-1, splashing type-2, and canopy, and further examine crown anomalies, including canopy formation and hole development, across a range of parameters. A regime map in the $A_r$–$\We$ space is developed to capture the transitions between these distinct behaviors. We further analyze how crown dynamics initiates interfacial instabilities that lead to node formation and finger growth, employing a novel simulation-based data extraction technique combined with linear stability analysis to estimate the number of fingers under different conditions. Below, we present the various splashing morphologies, followed by a detailed linear stability analysis.

\subsection{Splashing morphology}

\begin{figure}
\centering
\includegraphics[scale=0.5]{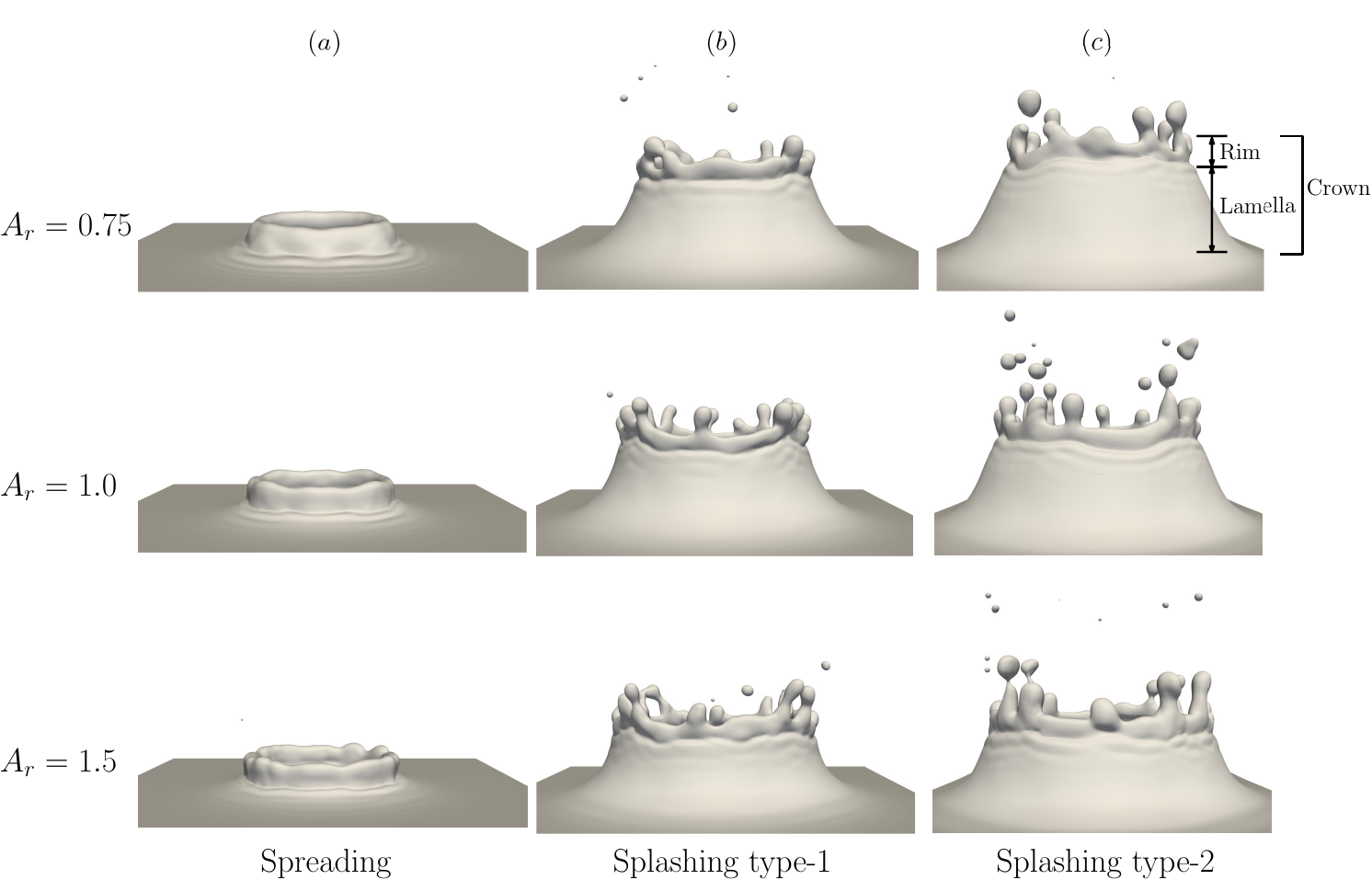}
\caption{Illustration of the different splashing behaviours for droplets with varying aspect ratios at (a) $\We = 117$, (b) $\We = 467$, and (c) $\We = 729$. Representative time instances are selected to highlight the crown–lamella evolution in spreading, splashing type-1, and splashing type-2 regimes. Here, $h = 2R_{eq}$, and the fluid system corresponds to a water–air system.}
\label{fig:splashing_regimes}
\end{figure}

We begin the presentation of our results by demonstrating the splashing morphologies exhibited by prolate ($A_r = 0.75$), spherical ($A_r = 1$), and oblate ($A_r = 1.5$) drops upon impacting a liquid film. Figure~\ref{fig:splashing_regimes}$(a–c)$ presents representative cases illustrating the distinct splashing behaviors observed for $A_r = 0.75$, 1.0, and 1.5 at $\We = 117$ ($\tau = 9.52$), $\We = 467$ ($\tau = 28.57$), and $\We = 729$ ($\tau = 59.5$), respectively. It can be seen that, at $\We=117$, the spreading regime features a radially expanding crown with minimal rim undulations. As the aspect ratio increases, particularly in oblate drops, these undulations become more prominent as shown in figure~\ref{fig:splashing_regimes}($a$). At $\We = 467$ and $\We = 729$, two distinct splashing behaviours, splashing type-1 and splashing type-2, are observed, as shown in figure~\ref{fig:splashing_regimes}($b$) and ($c$), respectively. These splashing modes are governed by different underlying mechanisms, which are discussed below.

\begin{figure}
\centering
\includegraphics[scale=0.5]{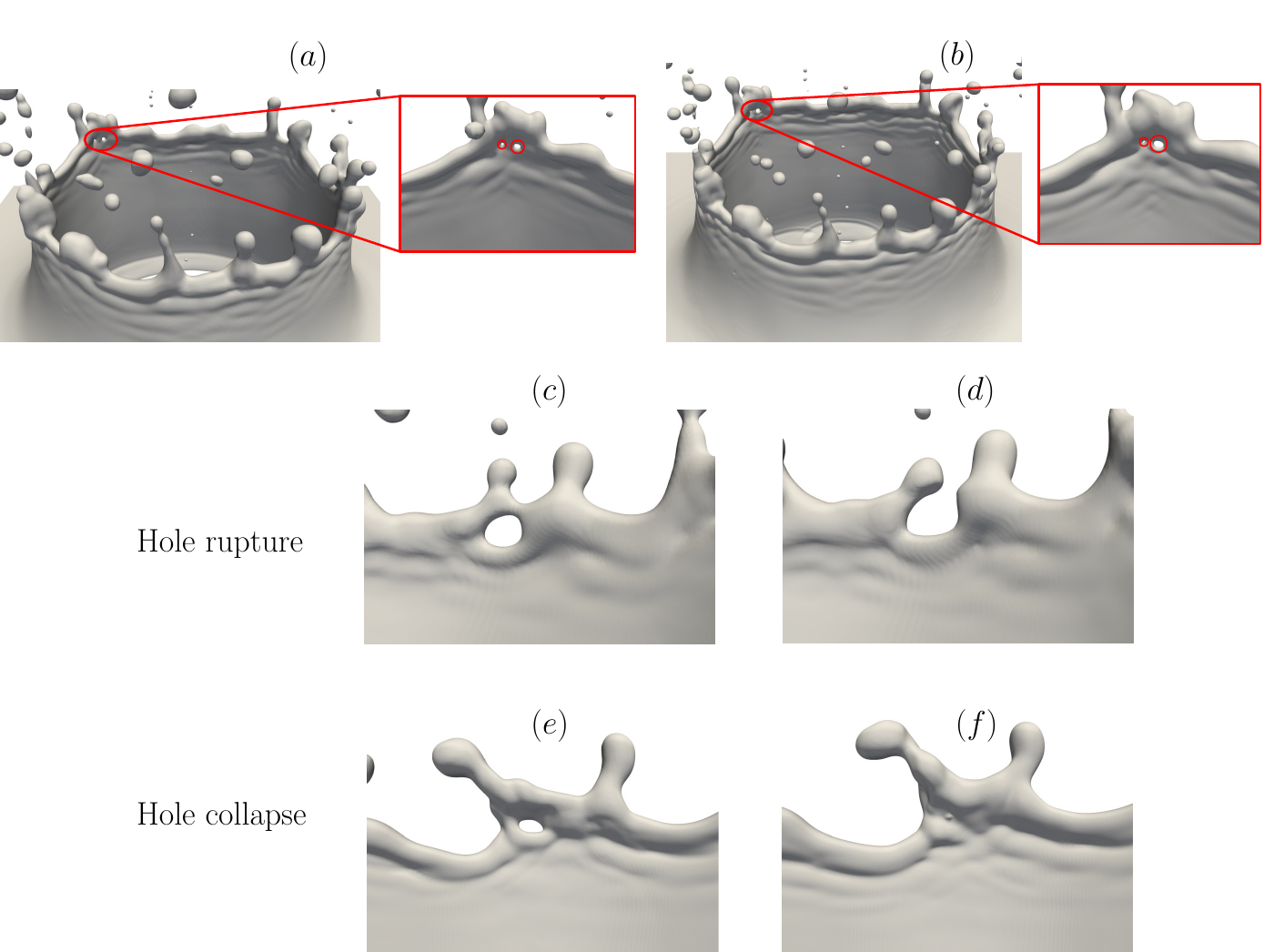}
\caption{Formation of holes due to crown sheet rupture leading to unstable splashing at $\tau = 35.2$ for $A_r = 2.0$ and $\We = 1050$ obtained using two mesh resolutions: ($a$) $\Delta x = 0.0149R_{eq}$ and ($b$) $\Delta x = 0.03R_{eq}$. Panels ($c$) and ($d$) depict the dynamics before and after hole rupture at $\tau = 22.61$ and $\tau = 25$, respectively, for $A_r = 2.0$ and $\We = 729$. Similarly, panels ($e$) and ($f$) show the crown dynamics before and after hole collapse at $\tau = 30.28$ and $\tau = 33.14$, respectively, for $A_r = 2.0$ and $\We = 1050$. Here, the fluid system considered is a water-air system.}
\label{fig:Hole_demonstration}
\end{figure}

\begin{figure}
\centering
Splashing type - 1 \hspace{2.8cm}  Splashing type - 2
\includegraphics[scale=0.5,trim={0 0 0 55pt}, clip]{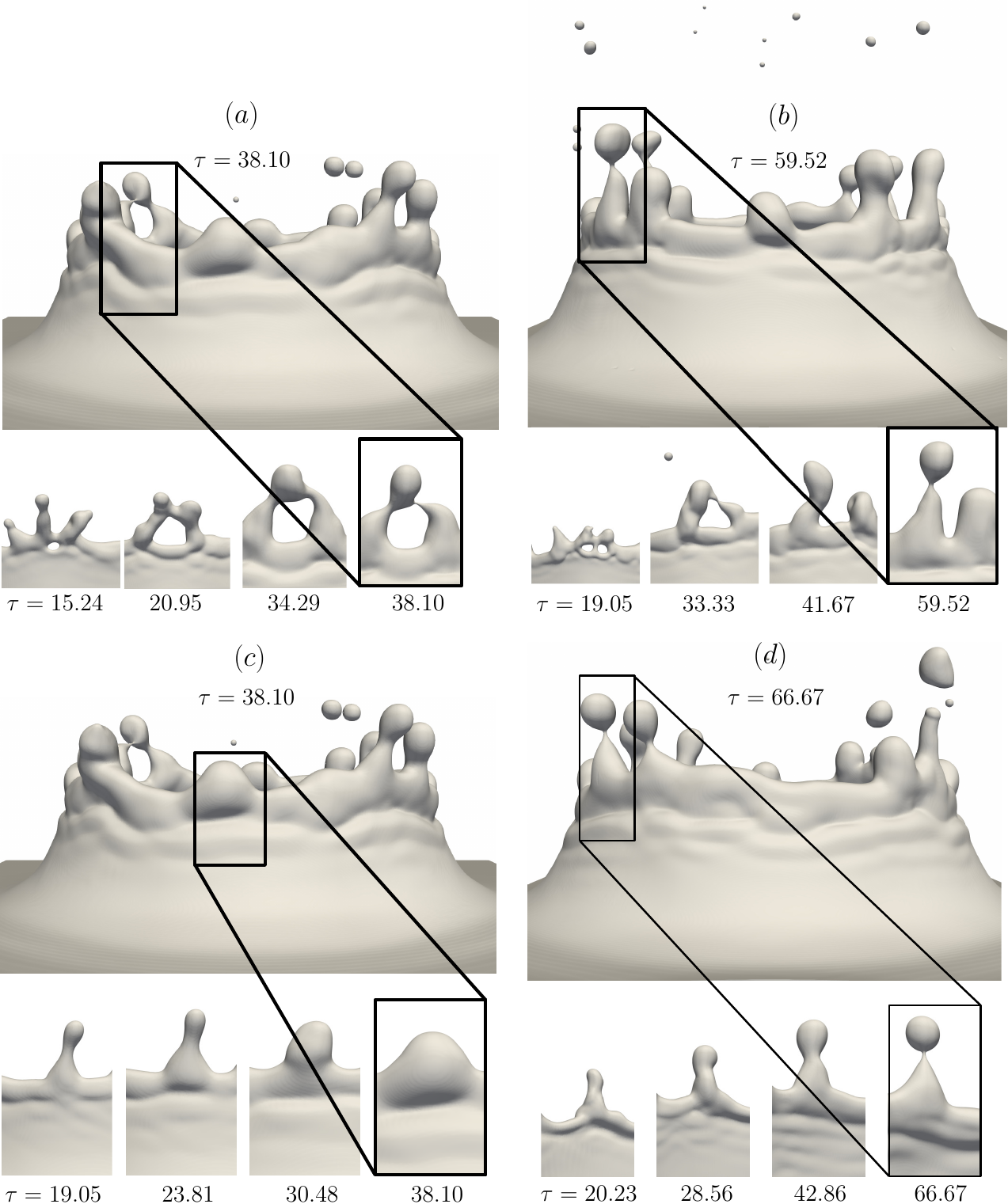}
\caption{Temporal evolution of the crown rim morphologies for a drop with $A_r = 1.5$, where panels ($a,c$) correspond to the splashing type-1 ($\We = 467$) and panels ($b,d$) represent the splashing type-2 ($\We = 729$).}
\label{fig:hole_finger_comp}
\end{figure}

A key feature common to all splashing impacts is the formation of holes in the crown lamella just beneath the crown rim, where the film is thinnest. These ruptures, referred to as holes, are observed across all aspect ratios; however, they become more pronounced for $A_r > 1$. Figure~\ref{fig:Hole_demonstration}($a$) illustrates this phenomenon for $A_r = 2.0$ and $\We = 1050$. The holes either collapse (figure~\ref{fig:Hole_demonstration}$e,f$) or result in finger splitting (figure~\ref{fig:Hole_demonstration}$c,d$), introducing variability in finger quantification and potential droplet detachment due to the momentum imparted onto the fingers by these ruptured holes. To confirm that these features are physical and not numerical artifacts, we have conducted additional simulations with significantly finer mesh resolution (figure~\ref{fig:Hole_demonstration}$b$). \cite{cossali2004role} reported that crown lamella thickness is non-uniform, typically ranging from $0.07R_{\text{eq}}$ to $0.12R_{\text{eq}}$ in experiments across various $\We$ and film thicknesses. Accordingly, the minimum grid size was selected to be well below this range, with adaptive mesh refinement applied sufficiently in advance of hole formation. The consistent appearance of holes at the same spatial locations and times in coarse and refined meshes confirms that these ruptures are physical phenomena and not artifacts of numerical resolution. These holes are likely triggered by undulations beneath the crown rim, which cause localised rupture in the thinning lamella. In contrast, other studies \citep{burzynski2018droplet, marston2016crown} have attributed similar ruptures to the bursting of air bubbles entrapped just before impact and traveling through the crown lamella; however, those rupture sites typically appear farther from the rim. Notably, our simulations show no evidence of such bubble entrapment prior to impact. Upon rupture, these holes generate jets that carry substantial momentum. The intensity of this jetting governs the breakup behaviour. When the momentum released during hole collapse is moderate, the jets thicken and merge into the rim without further detachment. In contrast, when the released momentum is high, the jets elongate and eventually pinch off to form droplets. Thus, hole rupture not only initiates fingers but also provides the momentum that drives droplet ejection, making hole-induced detachment a common feature of the splashing regime. Additional breakup may occur if other instabilities act on the rim. The distinction between these instabilities is discussed next in figure~\ref{fig:hole_finger_comp}.

Figure \ref{fig:hole_finger_comp} compares the two splashing regimes that emerge from the interplay of hole-driven droplet detachment and rim deceleration dynamics. In both splashing regimes, i.e., splashing type 1 and type 2, the collapse of holes beneath the rim produces fingers that can detach droplets. This hole-induced mechanism is therefore common to both type 1 and type 2 splashing. The difference lies in the behaviour of the crown rim itself. As the crown expands radially, the rim experiences momentum loss, causing the rim of the crown to decelerate. This deceleration makes the rim susceptible to Rayleigh-Taylor (RT) instability. In the splashing type-1 regime ($\We = 467$, figure\ref{fig:hole_finger_comp}$a$), the rim deceleration is weak; the RT growth rate remains low, and perturbations generated along the rim decay without causing further breakup. In contrast, in the splashing type-2 regime ($\We = 729$, figure\ref{fig:hole_finger_comp}$b$), the rim undergoes much stronger deceleration, amplifying RT perturbations and producing pronounced rim undulations that evolve into elongated fingers. These fingers elongate vertically and detach near their tips, forming secondary droplets. Thus, while both regimes exhibit hole-induced jet ejection, only type 2 shows rim deceleration-driven detachment.

\begin{figure}
\centering
\includegraphics[scale=0.5]{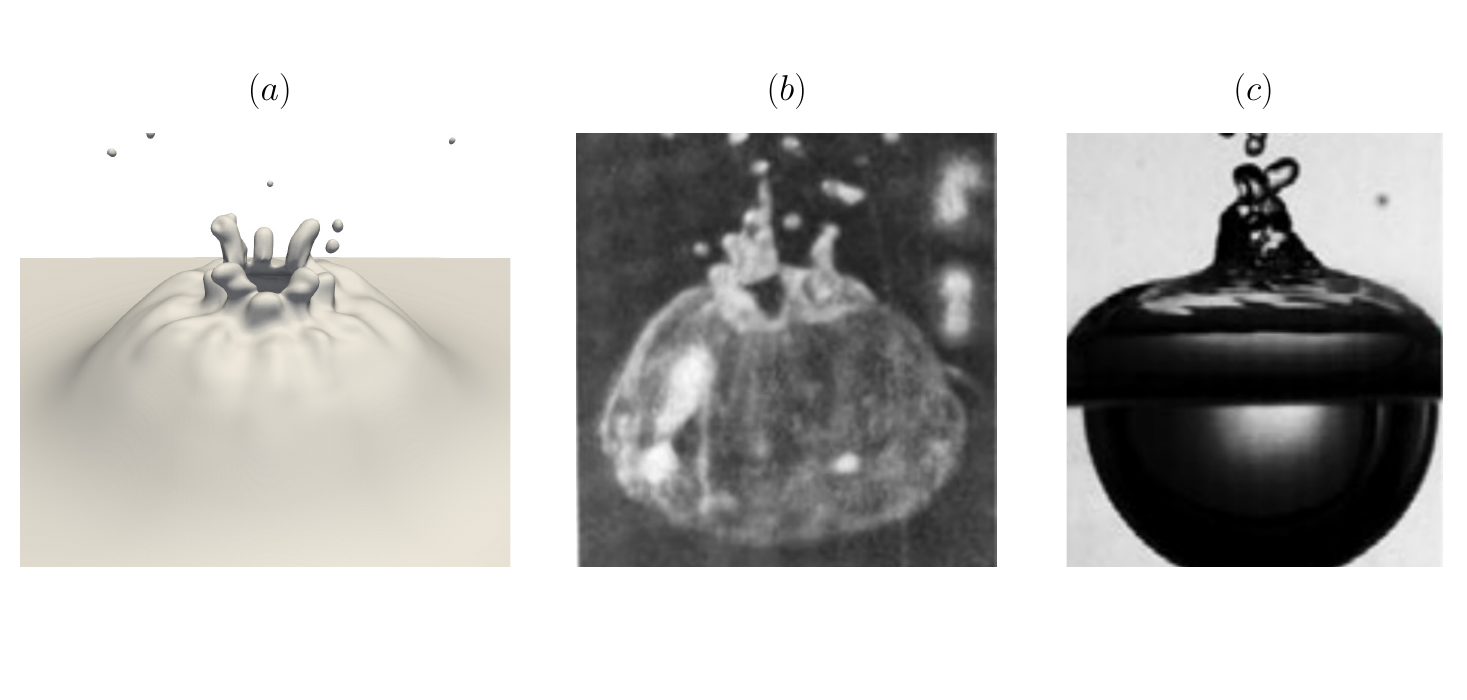}
\vspace{-1.0cm}
\caption{$(a)$ Formation of a canopy is observed for $A_r = 0.5$, $\We = 1050$. Similar crown structures have been reported previously $(b)$ by \cite{worthington1908study} for a water-diluted milk droplet impacting a water surface at $\We \approx 1430$, and $(c)$ by \cite{dandekar2025splash} in an air–water system for a droplet impacting a water pool at $\We = 850$.}
\label{fig:Inward_falling_crown}
\end{figure}

Another striking phenomenon observed during drop impact is canopy formation, which occurs when the vertical growth rate of the crown surpasses its radial expansion. This leads to the emergence of a hollow, inward-falling canopy structure, as shown in figure~\ref{fig:Inward_falling_crown}. At higher impact velocities, the crown rises rapidly, with the rim directed upward rather than outward. While Rayleigh–Taylor (RT) instabilities along the rim continue to initiate cusp formation and subsequent finger development, the breakup of these fingers into secondary droplets continues to be governed by Rayleigh–Plateau (RP) instability. Unlike the classical outward-expanding crown, the inward collapse in the canopy regime significantly alters the fragmentation dynamics. Canopy formation was first documented by \cite{worthington1908study} and has since been reported by several researchers, including \cite{dandekar2025splash} and \cite{wang2023analysis}, primarily in the context of spherical droplet impacts on deep liquid pools, where the unconfined vertical space permits substantial crown rise. In contrast, our simulations reveal canopy formation even on a liquid film when the impacting drop is prolate. The slender geometry of prolate drops results in a smaller initial contact area and narrower crown base diameter ($D_{crown}$), favoring rapid vertical expansion over radial expansion. Similar behavior was noted by \cite{deka2017regime} and \cite{thoraval2016vortex}, who showed that prolate drops are more prone to canopy formation than spherical or oblate drops, particularly in horizontally confined geometries at lower Weber numbers ($\We \approx 30$–100). At these low values of $\We$, the resulting canopy exhibits minimal rim undulations and lacks pronounced cusp development, leading to what is often termed a spreading canopy. \citet{eshraghi2020seal} further showed that canopy (splash-curtain) closure is strongly influenced not only by surface tension but also by the pressure difference between the air and the cavity, and they linked this closure to subsequent jet formation. Their study concluded that the key parameter governing canopy formation is the velocity of the air rushing into the cavity, rather than the impact velocity itself.

\begin{figure}
\centering
$(a)$ \\
\includegraphics[scale=0.65]{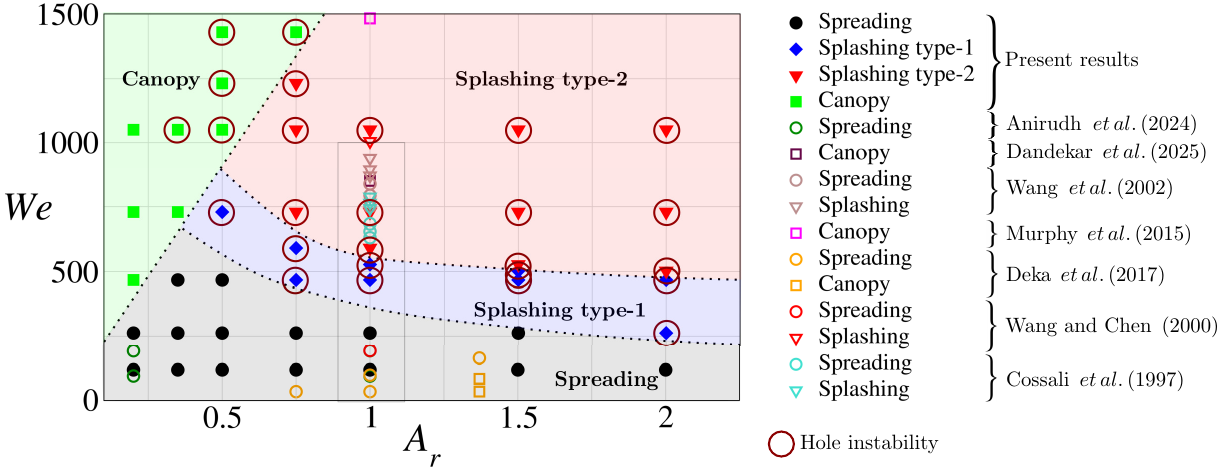} \\
\vspace{5mm}
 \hspace{0cm}  $(b)$ \hspace{2.5cm} $(c)$  \hspace{2.5cm} $(d)$  \hspace{2.5cm} $(e)$ \\
\includegraphics[scale=0.32]{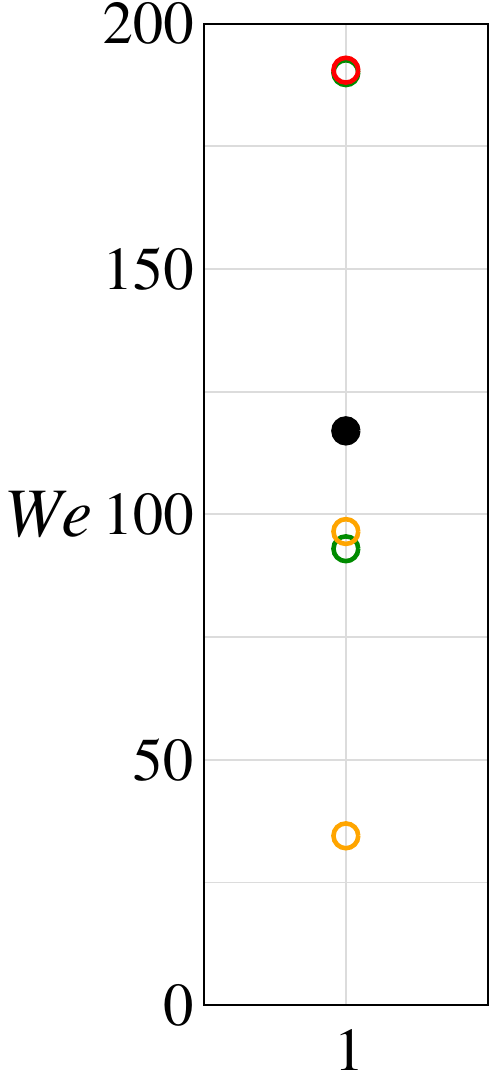}  \hspace{0.8cm}
\includegraphics[scale=0.32]{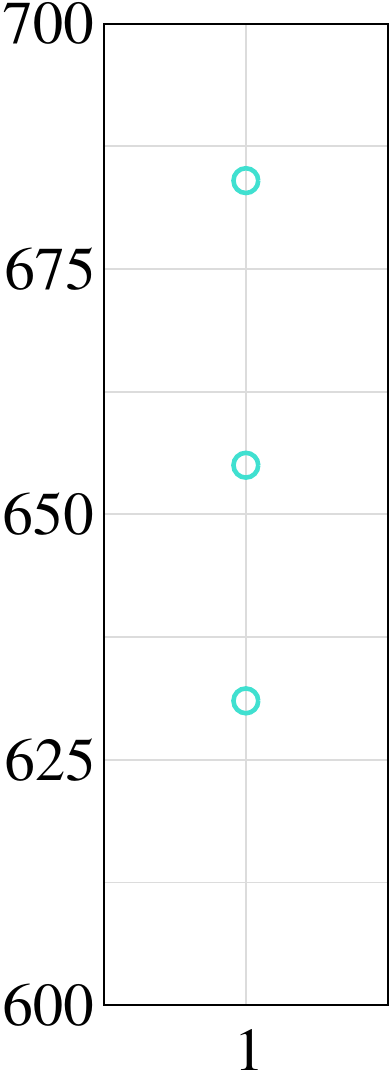} \hspace{0.8cm}
\includegraphics[scale=0.32]{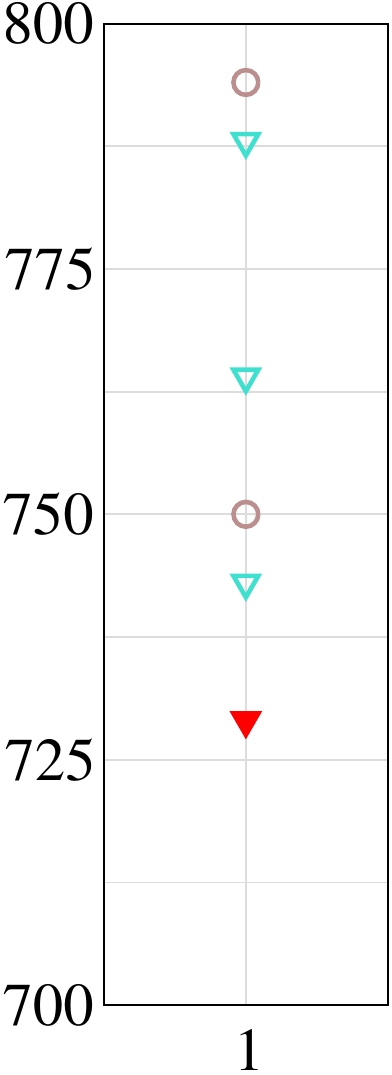} \hspace{0.8cm}
\includegraphics[scale=0.32]{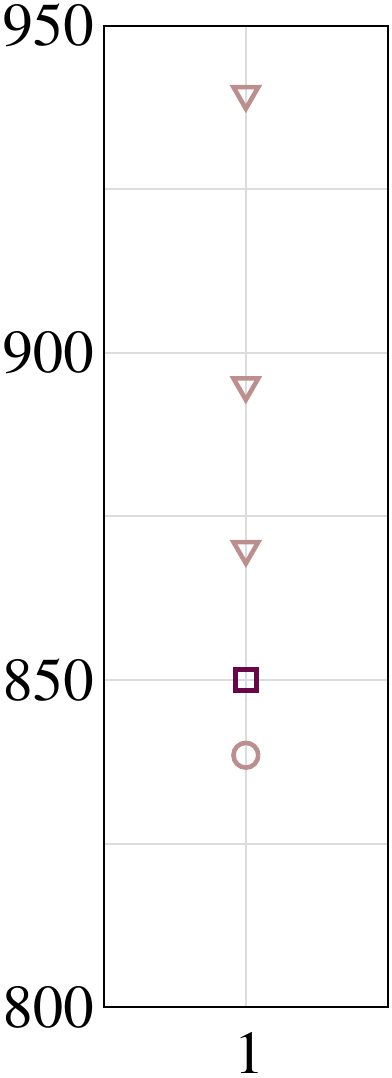} \hspace{0.8cm}
\caption{Regime map demarcating the coalescence dynamics and splashing morphology of a non-spherical drop impacting a liquid pool for various values of aspect ratio ($A_r$) and Weber number ($\We$), with $h = 2.0$. The regime map also includes results from prior studies for reference and comparison. The parameters and configurations considered in previous studies are summarized as follows. \citet{wang2002some}: $h = 2.0$, $A_r=1$; \citet{cossali1997impact}: $h = 1.4$, $A_r=1$; and \citet{wang2000splashing}: $h = 1.0$, $A_r=1$. Note that \citet{anirudh2024coalescence} and \citet{deka2017regime} examined the impact of non-spherical water droplets on a liquid–air interface, and \citet{murphy2015splash} investigated the impact of spherical droplets on an oil–water film. Panels $(b–e)$ present magnified views for $A_r = 1$.}
\label{fig:Regime_map}
\end{figure}

The regime map depicted in figure~\ref{fig:Regime_map} delineates the coalescence dynamics and splashing morphology of a non-spherical drop impacting a liquid film as a function of the aspect ratio ($A_r$) and Weber number ($\We$). The film thickness is fixed at $h = 2.0$ in our simulations. We perform simulations for discrete values of aspect ratio $A_r = 0.2$, 0.35, 0.5, 0.75, 1.0, 1.5, and 2.0 across Weber numbers $\We = 117$, 262, 467, 729, and 1050. To ensure complete coverage of the regime map, additional simulations are carried out for $A_r = 1.5$ and 2.0 at $\We = 502$; $A_r = 1.0$ and 1.5 at $\We = 527$; $A_r = 0.75$ and 1.0 at $\We = 591$; and $A_r = 0.5$ and 0.75 at $\We = 1232$ and 1429. The regime map (figure~\ref{fig:Regime_map}) also incorporates results from previous studies for reference and comparison. Among these, while \cite{anirudh2024coalescence} and \cite{deka2017regime} investigated the impact of non-spherical water drops on a liquid-air interface, the other earlier studies focused exclusively on spherical drops. Notably, \cite{deka2017regime} examined the dynamics of a non-spherical drop impacting a restricted pool and reported that canopy structures form more readily in prolate drops than in oblate drops, an observation that is similar to that observed in the present study. In addition, \cite{wang2000splashing, wang2002some, cossali1997impact} explored the influence of film thickness ($h$) on splashing dynamics, albeit for spherical drops ($A_r = 1$). To maintain visual clarity, magnified views of the regime map are shown in figure~\ref{fig:Regime_map}($b-e$) for $A_r = 1.0$.

In figure~\ref{fig:Regime_map}, we observe that at $\We = 117$, droplets of all aspect ratios undergo complete spreading, merging smoothly with the liquid film without generating rim instabilities or secondary structures. As $\We$ increases, secondary droplets begin to detach from fingers through hole rupture (splashing type-1) for $A_r>0.5$. At $\We = 467$, drops with $A_r = 0.75$, 1.0, 1.5, and 2.0 display splashing type-1, while those with $A_r = 0.35$ and 0.5 continue to spread.  At this Weber number, the drop with $A_r = 0.2$ transitions to the canopy regime. At $\We = 729$, splashing type-2 emerges for $A_r \geq 0.75$, whereas $A_r = 0.5$ remains in splashing type-1, and $A_r = 0.35$ and 0.2 show canopy formation. Additional simulations with smaller $\We$ increments in the range $467 < \We < 729$ reveal that higher aspect ratios promote stronger finger growth, eventually leading to splashing type-2. At $\We = 1050$, canopy structures appear for $A_r = 0.5$, 0.35, and 0.2. The canopy region increases as we increase the $\We$. At higher Weber numbers, canopy formation persists at $\We = 1232$ for $A_r = 0.5$ but is absent for $A_r = 0.75$. In contrast, at $\We = 1429$, both $A_r = 0.5$ and 0.75 exhibit canopy behavior. \citet{khan2024impact} experimentally showed that oblate drops are more susceptible to splash-induced fragmentation when impacting a water film, consistent with our observation that splashing becomes more pronounced at larger $A_r$. The dotted lines in figure~\ref{fig:Regime_map} represent notional boundaries separating the different regimes, highlighting the coupled influence of drop shape and impact inertia on the resulting dynamics.

\begin{figure}
\centering
\includegraphics[scale=0.5]{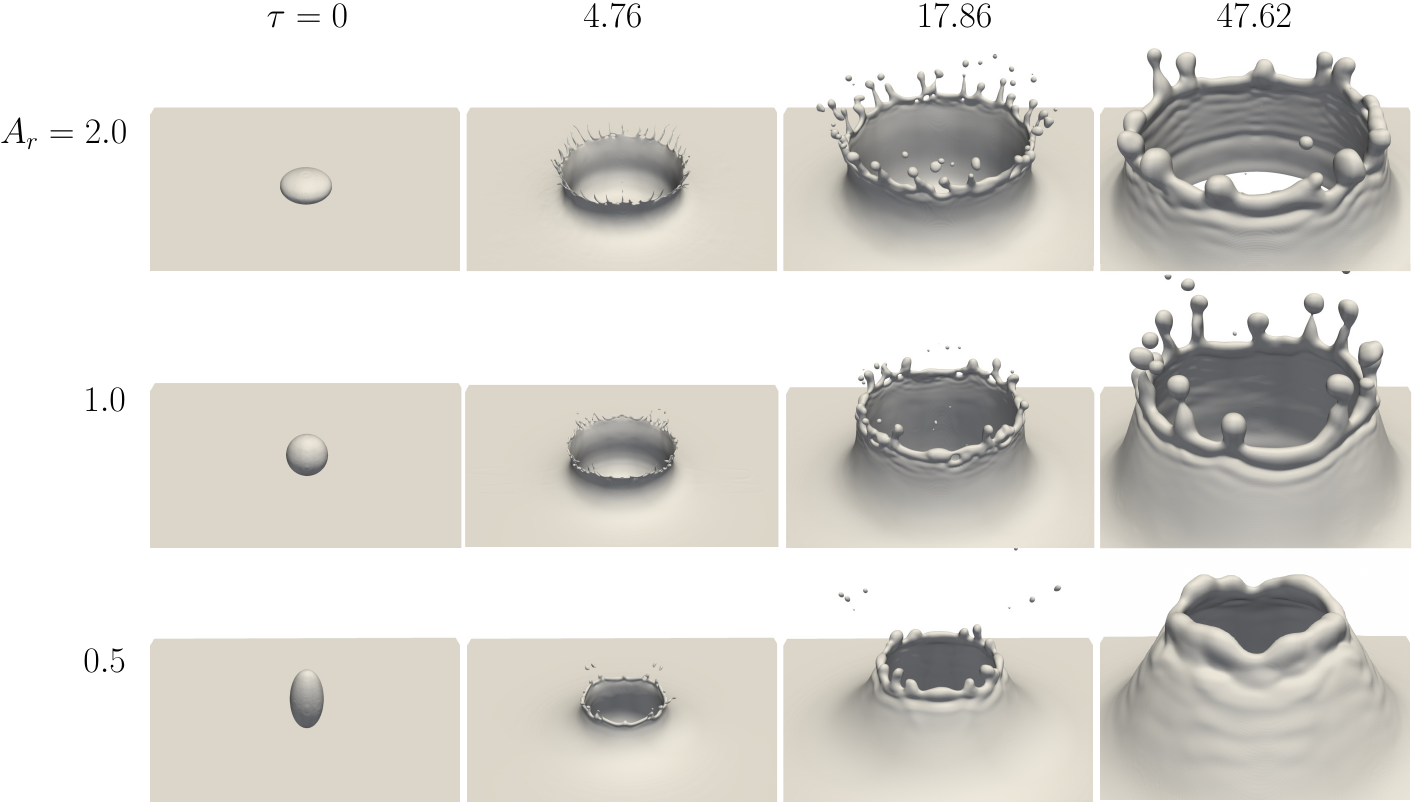}
\caption{Temporal evolution of the splashing dynamics for droplets with aspect ratios $A_r = 2.0$ (top panel), 1.0 (middle panel), and 0.5 (bottom panel), impacting a liquid film of height $h = 2$ at $\We = 729$.}
\label{fig:splashing_morph}
\end{figure}

To gain physical insight into the observation that splashing becomes more prominent for drops with higher aspect ratios ($A_r$), figure~\ref{fig:splashing_morph} presents the temporal evolution of splashing dynamics for drops with $A_r = 2.0$, 1.0, and 0.5 at $\We = 729$. At $\tau = 0$, the drop contacts the liquid film, initiating radial crown expansion. It can be seen that due to morphological differences, the oblate drop ($A_r = 2.0$) forms a broader crown than the prolate drop ($A_r = 0.5$), as evident at $\tau = 4.76$ in figure~\ref{fig:splashing_morph}. As the crown expands, fluid accumulates at the rim, leading to thickening. The rim acts like a cylindrical jet and becomes unstable due to the influence of surface tension, triggering undulations via the Rayleigh–Plateau (RP) instability. The narrower crown of the prolate drop decelerates more slowly than the oblate and spherical drops, resulting in less fluid accumulation at the rim and yielding a thinner structure. This, in turn, affects the RP instability by producing shorter azimuthal wavelengths for prolate drops, as compared to their oblate and spherical counterparts. By $\tau = 17.86$, the rim undulations become pronounced (mostly fully developed). Subsequently, as rim deceleration intensifies, the Rayleigh–Taylor (RT) instability amplifies these undulations into fingers, which are clearly visible in figure.~\ref{fig:splashing_morph}. Since rim deceleration is slower for prolate drops, the RT instability grows more slowly, delaying finger evolution. This pattern holds across all prolate drops ($A_r < 1.0$), which consistently exhibit delayed fingering compared to spherical and oblate drops. For $A_r = 0.5$, finger formation occurs much later than shown and is not captured within the current visualization timeframe. As seen in figure~\ref{fig:splashing_morph}, at $\tau = 47.62$, fingers are well developed in oblate and spherical drops, while those in prolate drops remain subdued, indicating reduced instability growth.

The temporal evolution of splashing reveals morphological differences across various non-spherical drop shapes. However, it does not provide a quantitative understanding of the associated mechanisms responsible for node formation and finger growth. To address this, we employ linear stability analysis to identify the instabilities that drive node initiation and development induced by crown geometry. The governing parameters for the stability analysis, such as rim radius, rim deceleration, and crown diameter, are accurately extracted from our numerical simulations and incorporated into the linear stability framework to predict the number of nodes that evolve into fingers and their corresponding growth rates. The details of the linear stability analysis are presented in the next section.

\subsection{Linear stability analysis} \label{sec: LSA}

\begin{figure}
\centering
\includegraphics[scale=0.7]{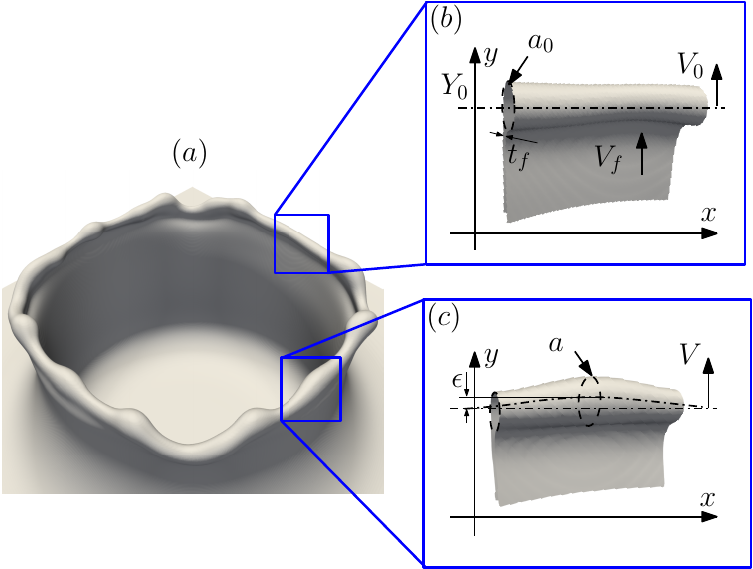}
\caption{$(a)$ A typical crown obtained from numerical simulations. $(b)$ The unperturbed base state of the rim, where $t_f$ denotes the crown lamella thickness and $V_f$ is the velocity of the crown lamella. $(c)$ The perturbed state of the rim (base state plus disturbance). Here, $a$ represents the cross-sectional area of the cylindrical structure resting on a thin liquid sheet, and $V$ is the rim velocity. The subscript 0 indicates quantities associated with the unperturbed base state, whereas perturbed quantities used in the linear stability analysis are denoted without subscripts.}
\label{fig:LSA_schematic}
\end{figure}

The linear stability analysis examines the behavior of an infinitesimally small disturbance imposed on the rim of the crown. Figure \ref{fig:LSA_schematic}(a–c) illustrates the typical crown shape obtained from numerical simulation, the unperturbed (base state) and perturbed (base state plus disturbance) cross-sections of the rim, respectively. To examine the base state and its stability characteristics, we employ a Cartesian coordinate system $(x, y)$, with the rim centerline located at $y = Y_0(t)$, as shown in figure \ref{fig:LSA_schematic}$(b)$. The time-dependent cross-sectional area of the rim is given by $a_0 = \pi {r_0}^2$, where $r_0$ is the cross-sectional radius of the rim. The vertical rise velocity of the rim can be represented as $V_0 = dY_0/dt$. 

\subsubsection{Base state}
We model the unperturbed crown rim as a cylindrical structure resting at the edge of the sheet formed due to capillary forces  \citep{yarin1995impact, roisman2006spray}. The rim is treated as a quasi-one-dimensional liquid structure that grows due to mass influx from the lamella. Accordingly, the evolution of its cross-sectional area is governed by the mass conservation equation at $y=Y_0$ and can be expressed as:
\begin{equation}
\frac{d a_0}{d t} = t_f(V_f - V_0),
\label{conti}
\end{equation}
where $t_f$ denotes the crown lamella thickness, $V_f$ is the velocity within the crown lamella at $y=Y_0$, and $V_0$ represents the upward velocity of the rim. Furthermore, the rim dynamics are governed by the balance between inertia, capillary pressure, and the momentum flux from the crown lamella, leading to the momentum balance equation:
\begin{equation}
a_0 \frac{d {V_0}}{d t} = t_f(V_f - V_0)^2 - \frac{2\sigma}{\rho},
\label{eq 3.3}
\end{equation}
where $\sigma$ is the surface tension and $\rho$ is the liquid density.

\subsubsection{Perturbation equations}

To perform the linear stability analysis, we perturb the base state with an infinitesimally small transverse disturbance. The evolution of the centerline of the resulting perturbed rim, $Y(x,t)$, is then analyzed. This imposed transverse disturbance induces a velocity component $u$ in the $x$-direction within the rim, which is much smaller than the vertical velocity $V$. As a result, the total velocity field is expressed as
\begin{equation}
 v = u \mathbf{e}_x + V \mathbf{e}_y,
 \end{equation}
where $\mathbf{e}_x$ and $\mathbf{e}_y$ are the unit vectors in the $x$ and $y$ directions, respectively. In the linear stability framework, the wavelength of the long-wave perturbation $(\lambda)$ satisfies
\begin{equation}
| Y - Y_0 | = \epsilon \ll \lambda \quad \text{and} \quad r_0 \ll \lambda,
\end{equation}
indicating that the perturbation amplitude $\epsilon$ is much smaller than its wavelength. This also ensures that
\begin{equation}
\left| \frac{\partial Y}{\partial x} \right| \ll 1.
\end{equation}
Under this approximation, the $x$-component of the velocity $u$, the velocity gradient $\partial V / \partial x$, and the gradient of the angular velocity of the rim cross-section $\partial \Omega / \partial x$ are assumed to be sufficiently small, rendering all nonlinear convective terms in the momentum equation second-order in magnitude, and they are neglected in the linear stability analysis \citep{entov1984dynamics}. Here, $\Omega = {\partial V/\partial x}$. The resulting linearized perturbation equations are given by
\begin{subequations}
\begin{equation} 
\frac{\partial a}{\partial t} + {a_0} \frac{\partial u}{\partial x} - {t_f}( V_f - V) = 0, \label{dim_mass_con}
\end{equation}
\begin{equation}
\rho {a_0} \frac{\partial u}{\partial t} - \frac{\partial P}{\partial x} - \left[ - \rho {t_f} (V_f -V) u + 2 \sigma \frac{\partial Y}{\partial x} \right ] = 0,
\label{dim_xmom_con}
\end{equation}
\begin{equation}
\rho a \frac{\partial V}{\partial t} - P \kappa - \frac{\partial Q}{\partial x} - \left[ \rho {t_f} (V_f - V)^2 - 2\sigma \right] = 0,
\label{dim_ymom_con}
\end{equation}
\begin{equation}
\rho \frac{\partial L}{\partial t} - \frac{\partial M}{\partial x} - Q - {m_S} = 0. \label{dim_amom_con}
\end{equation}
\end{subequations}
Here, $\kappa$ denotes the curvature of the centerline of the rim, $Q$ is the shear force in the $y$-direction acting on the cross-section of the rim, and $m_S$ represents the distributed moment due to external forces. The magnitude of the longitudinal tensile force ($P$), which accounts for both the capillary pressure within the rim and the contribution from surface tension, is given by
\begin{equation}
P = \pi \sigma r + \sigma {a_0} \frac{\partial^2 r}{\partial x^2}.
\end{equation}
The angular momentum of the rim per unit length in the plane ($L$) is given by
\begin{equation}
L = I_0 \Omega, \quad \text{where} \quad I_0 = \frac{\pi {r_0}^4}{4}.
\end{equation}
The internal moment resulting from stresses within the cross-section ($M$), produced by the pressure gradient in the $y$-direction in the rim cross-section due to its acceleration, is given by
\begin{equation}
M = -\rho I_0 \frac{d V}{d t}.
\end{equation}
The moment due to the flow from the crown lamella into the rim is given by
\begin{equation}
m_S = -\rho {t_f} {r_0} (V_f - V_0) u.
\end{equation}

\subsubsection{Normal mode approximation}

We investigate the temporal stability of the base state to infinitesimal two-dimensional (2D) disturbances by employing a normal mode approximation. Here, each flow variable is expressed as the sum of the base state and a 2D perturbation, which is further decomposed into an amplitude and a wave-like component as follows:
\begin{equation}
(Y,r,u,Q)(x,y,t) = [Y_0, {r_0}, 0, 0]+ \left(\hat{\epsilon}, \hat{\eta}, \hat{u}, \hat{q} \right) \exp(\omega_c t + \i \alpha x), \label{perturbations}
\end{equation}
where, the hats denote the perturbation quantities, $\alpha = 2 \pi/\lambda$ is the (real) wavenumber, and $\omega_c$ is the complex frequency, with $\omega = \text{Real}(\omega_c)$ representing the growth rate of the disturbance. For disturbances of the form given in eq. (\ref{perturbations}), $\omega > 0$ corresponds to a temporally unstable mode, $\omega < 0$ indicates a stable mode, and $\omega = 0$ denotes a neutrally stable disturbance. Note that the quantities $r_0$, $a_0$, and $I_0$ of the rim vary with time; however, their temporal evolution is assumed to be sufficiently slow compared to that of the disturbances. Hence, a quasi-steady-state approach is adopted by perturbing the base state at a frozen time. Following the standard procedure of linear stability analysis, i.e. substituting eq.~(\ref{perturbations}) into the governing equations, subtracting the base state equations, and subsequently linearizing the resultant equations, we obtain the governing dispersion equation for the linear stability analysis. In dimensionless form, the dispersion relation for the growth rate of the disturbance can be written as:
\begin{equation}
\begin{aligned}
    & 2(2 + {\bar \alpha}^2)\pi^2 {\bar \omega}^4 + 4(3 + {\bar \alpha}^2)\pi {\bar W}_0 {\bar \omega}^3 \\
    & + \left[ 2(4 + {\bar \alpha}^2) {\bar W}_0^2 - 2\pi {\bar t_f} (2 + {\bar \alpha}^2) {d {\bar V_0} \over dt} + \pi^2 {\bar \alpha}^2 (2 + {\bar \alpha}^2 + {\bar \alpha}^4) \right] {\bar \omega}^2 \\
    & + \left[ 2 {\bar \alpha}^2 \left (4 - {\bar t_f} - {\bar t_f} {d {\bar V_0} \over dt} \right) - 4 {\bar t_f} {d {\bar V_0} \over dt} + \pi {\bar \alpha}^6 + {\bar \alpha}^4 (3\pi + 2 {\bar t_f}) \right] {\bar W}_0 {\bar \omega} \\
    & - 2\pi^2 {\bar \alpha}^4 (1 - {\bar \alpha}^2) + 4\pi {\bar \alpha}^2 (2 + {\bar \alpha}^2){d  {\bar V_0} \over dt} = 0.
\end{aligned}\label{dispersion_eqn}
\end{equation}
Here,
\begin{equation}
\omega = \sqrt{ \sigma \over \rho {r_0}^3} {\bar \omega}, ~ \alpha = {{\bar \alpha} \over r_0}, ~ t_f = {\bar t_f} r_0, ~ {\dot V_0} ={\sigma \over \rho {r_0}^2} {\bar {\dot V}}, ~ {\rm and} ~ {\bar W}_0 = \rho t_f (V_f - V_0) / \sqrt{\rho \sigma r_0},
\end{equation}
wherein ${\dot V_0} = {d V_0 / d t}$ is the acceleration of the rim. The ‘bar’ superscript denotes dimensionless quantities, which are omitted in the subsequent analysis for clarity. The current stability analysis adopts the same methodology as that described in \cite{roisman2006spray,entov1984dynamics}.

\begin{figure}
\centering
\includegraphics[scale=0.7]{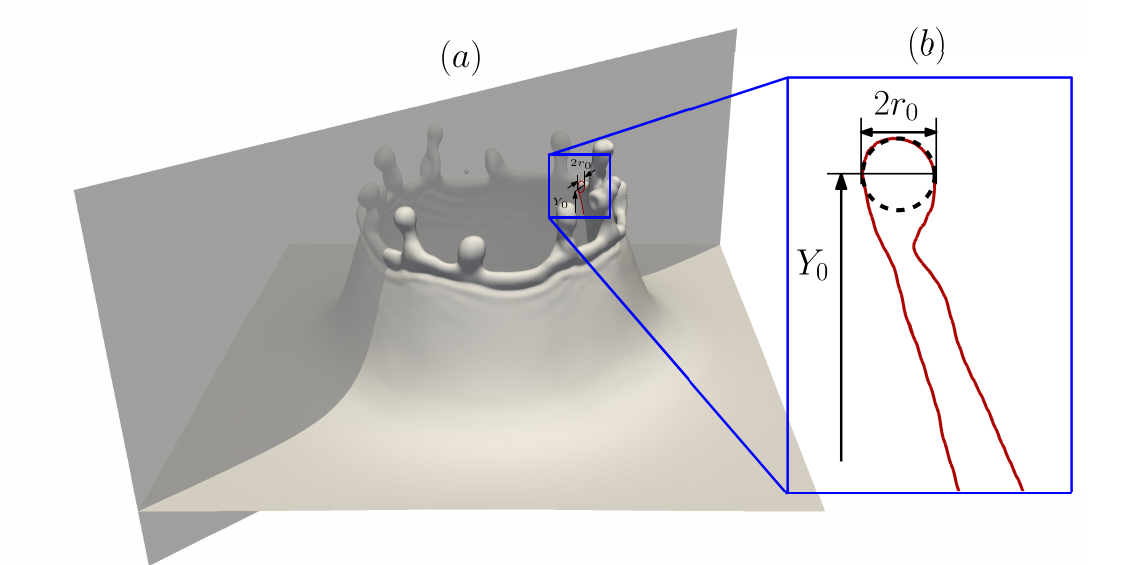}
\caption{($a$) Schematic showing the locations of slices used to extract the rim radius ($r_0$) and rim height ($Y_0$).
($b$) Illustration of the procedure for extracting $r_0$ and $Y_0$ from one of the slices shown in panel $a$.}
\label{fig:Data_extraction}
\end{figure}

The dispersion relation (eq. \ref{dispersion_eqn}) contains two unknown parameters, namely the rim radius ($r_0$) and the rim acceleration ($\dot{V}_0$). Previous studies on the linear stability of node formation during splashing have proposed various approximations for these parameters, each with inherent limitations. \citet{zhang2010wavelength} estimated $r_0$ by fitting circles to the inner and outer edges of the crown rim. Although this approach is effective during the initial stages of crown formation, it fails to represent the rim geometry accurately at later times when significant deformations occur. \citet{agbaglah2014growth} used axisymmetric numerical simulations to estimate $r_0$ and $\dot{V}_0$. However, since splashing is fundamentally an asymmetric phenomenon, the axisymmetric assumption reduces the accuracy and applicability of their results. Additionally, \citet{agbaglah2013longitudinal} conducted simulations on a crown segment within a computational domain spanning two perturbation wavelengths, introducing localized disturbances to investigate finger growth and droplet detachment. This approach does not capture the global temporal evolution of the crown or related phenomena such as hole formation, thus neglecting crucial aspects of rim dynamics. Using potential flow theory, \citet{bisighini2010crater} analytically modeled the evolution of a crater as a growing and translating sphere within a liquid pool. Their approach can, in principle, be extended to estimate crown dynamics, including $r_0$ and $\dot{V}_0$, using conservation laws, provided the crown thickness ($t_f$) is known. The present study focuses on impacts involving non-spherical drops, where the drop shape deviates significantly from a sphere. Extending the potential flow formulation to accommodate a growing and moving ellipsoid introduces substantial mathematical and computational challenges, making such an approach impractical for our purposes. Thus, in contrast to the previous approaches, we extract $r_0$ and $\dot{V}_0$ directly from numerical simulation results at the exact time when the number of nodes is to be determined. The rim radius is calculated by sampling data at different locations around the rim, as illustrated in figure~\ref{fig:Data_extraction}, and averaging the values to account for spatial variations. The rim acceleration is obtained by tracking the vertical position of the rim ($Y_0$) over time and applying a second-order central difference scheme to compute its second derivative. To minimize errors, we restrict the analysis to rim regions that have not yet developed fingers, as finger formation distorts the accuracy of both $r_0$ and $\dot{V}_0$. It is to be noted that the geometric approximations like those used by \citet{zhang2010wavelength}, which include the entire crown structure, including fingers, are prone to such inaccuracies. However, this approach allows the effects of hole rupture or collapse, the events that dynamically alter the rim structure, to be incorporated when extracting $r_0$ and $\dot{V}_0$. Following the convention of \citet{roisman2006spray}, the crown thickness ($t_f$) is assumed to be $0.5r_0$. Using these extracted parameters from the numerical simulations, we solve eq.~(\ref{dispersion_eqn}) to determine the growth rate ($\omega$) as a function of wavenumber ($\alpha$). From eq. \eqref{dispersion_eqn}, the dispersion relation depends solely on the rim deceleration, $\dot{V_0}$. The critical growth rate, $\omega_{cr}$, shows the strongest sensitivity to $\dot{V_0}$, whereas the corresponding wavenumber, $\alpha_{cr}$, is less sensitive. The rim deceleration obtained from the simulations is substituted into the dispersion relation to construct the dispersion curves, from which the maximum growth rate, $\omega_{cr}$, and the associated wavenumber, $\alpha_{cr}$, are determined. Using these quantities together with the rim radius $(r_0)$ and the crown diameter $(D_{crown})$, the number of fingers is estimated as $n = (\alpha_{cr}D_{crown}) /(2 r_0)$.

\subsection {Competition between RT and RP instabilities}

\begin{figure}
\centering
\hspace{1cm} $(a)$ \hspace{5cm} $(b)$ \\
\includegraphics[scale=0.18]{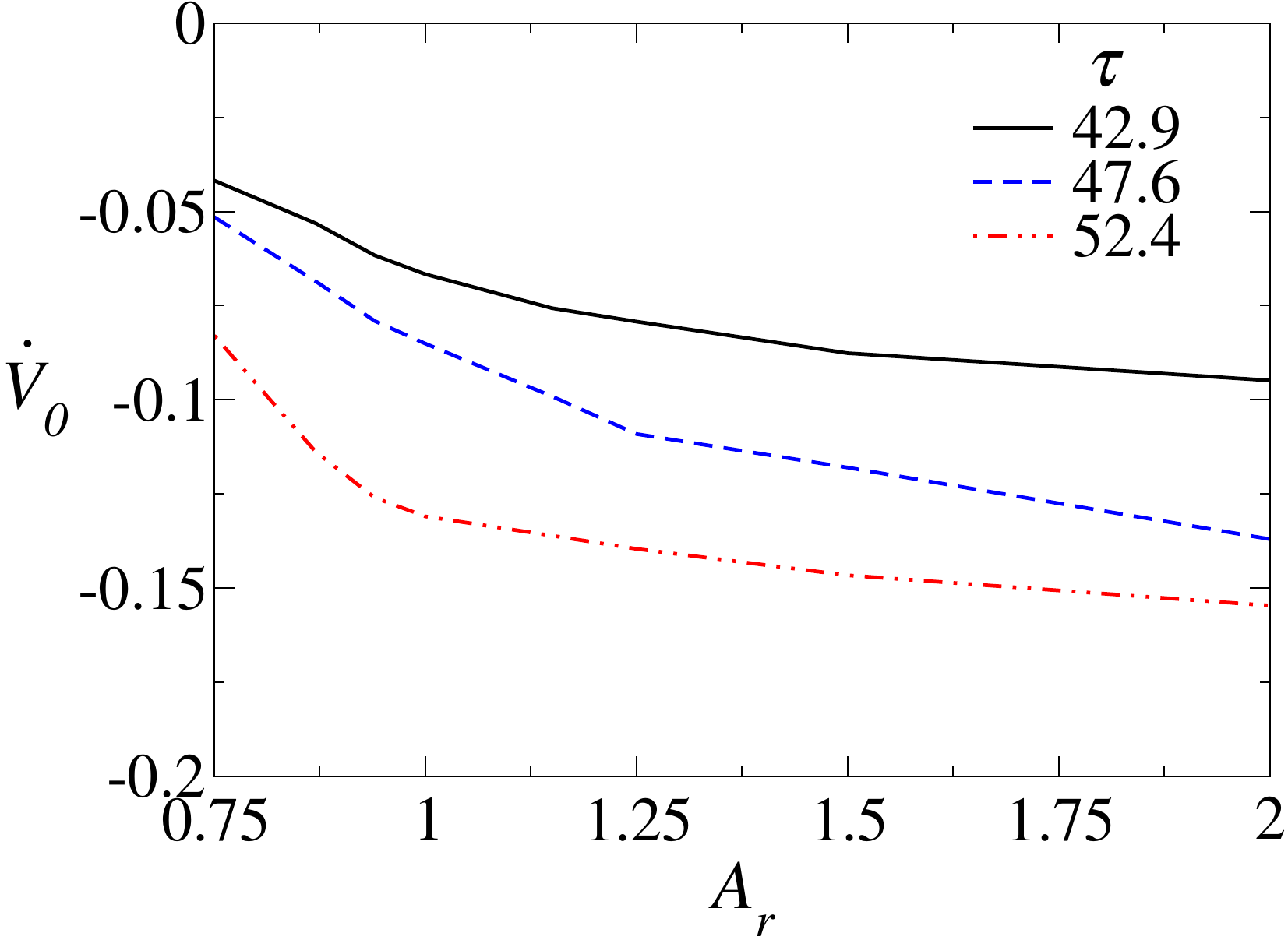}
\includegraphics[scale=0.18]{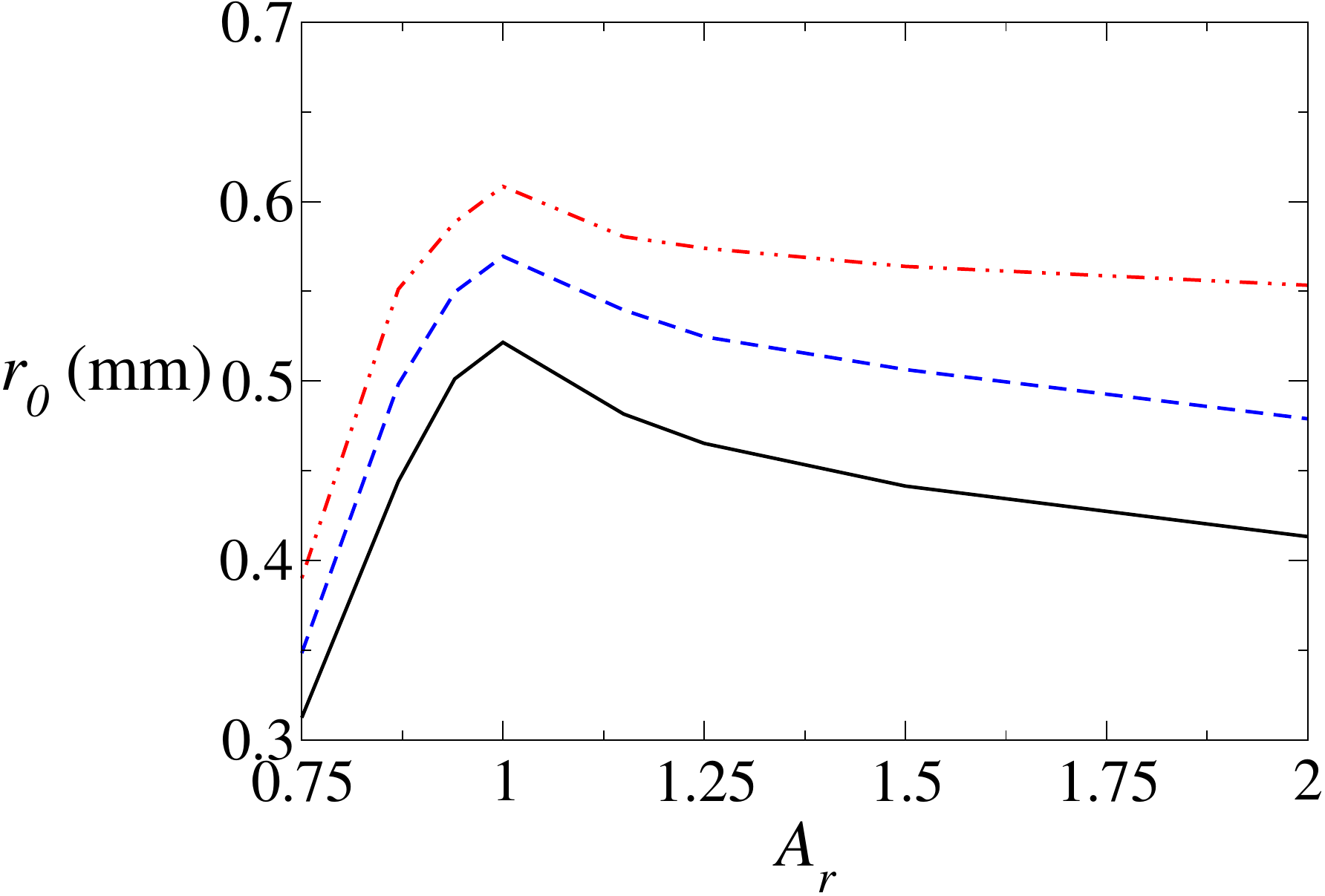}\\
\caption{($a$) Variation of rim deceleration $(\dot{V}_0)$ and ($b$) rim radius $(r_0)$ with aspect ratio $(A_r)$ for $We = 729$ at different time instances $\tau = 42.9$, 47.6, and 52.4.}
\label{fig:extracted_v}
\end{figure}

To further highlight the effect of aspect ratio, a general comparison of the input parameters ($r_0$, $\dot{V_0}$) with aspect ratio ($A_r$) is made at three different time-steps and is shown in figure \ref{fig:extracted_v}. The parameters $r_0$ and $\dot{V_0}$ exhibit similar trends with $A_r$ at different timesteps. As shown in figure \ref{fig:splashing_morph}, the rim deceleration $\dot{V_0}$ increases with $A_r$, such that prolate drops experience a smaller deceleration than oblate ones, consistent with figure \ref{fig:extracted_v}$(a)$. The rim radius $r_0$ is governed by two opposing effects: stronger deceleration promotes greater fluid accumulation and increases $r_0$, whereas a larger crown circumference associated with higher $A_r$ distributes the accumulated liquid over a wider perimeter, reducing $r_0$. The combined influence of these effects yields the trend in figure \ref{fig:extracted_v}$(b)$, where spherical drop exhibit the largest rim radius. These variations in $r_0$ and $\dot{V_0}$ directly influence the instability behaviour discussed earlier. The growth rate is primarily controlled by the rim deceleration, which increases with $\dot{V_0}$, while $r_0$ determines the dominant wavenumber and hence the number of nodes that develop along the rim.

\begin{figure}
\centering
\hspace{1cm} $(a)$ \hspace{5cm} $(b)$ \\
\includegraphics[scale=0.18]{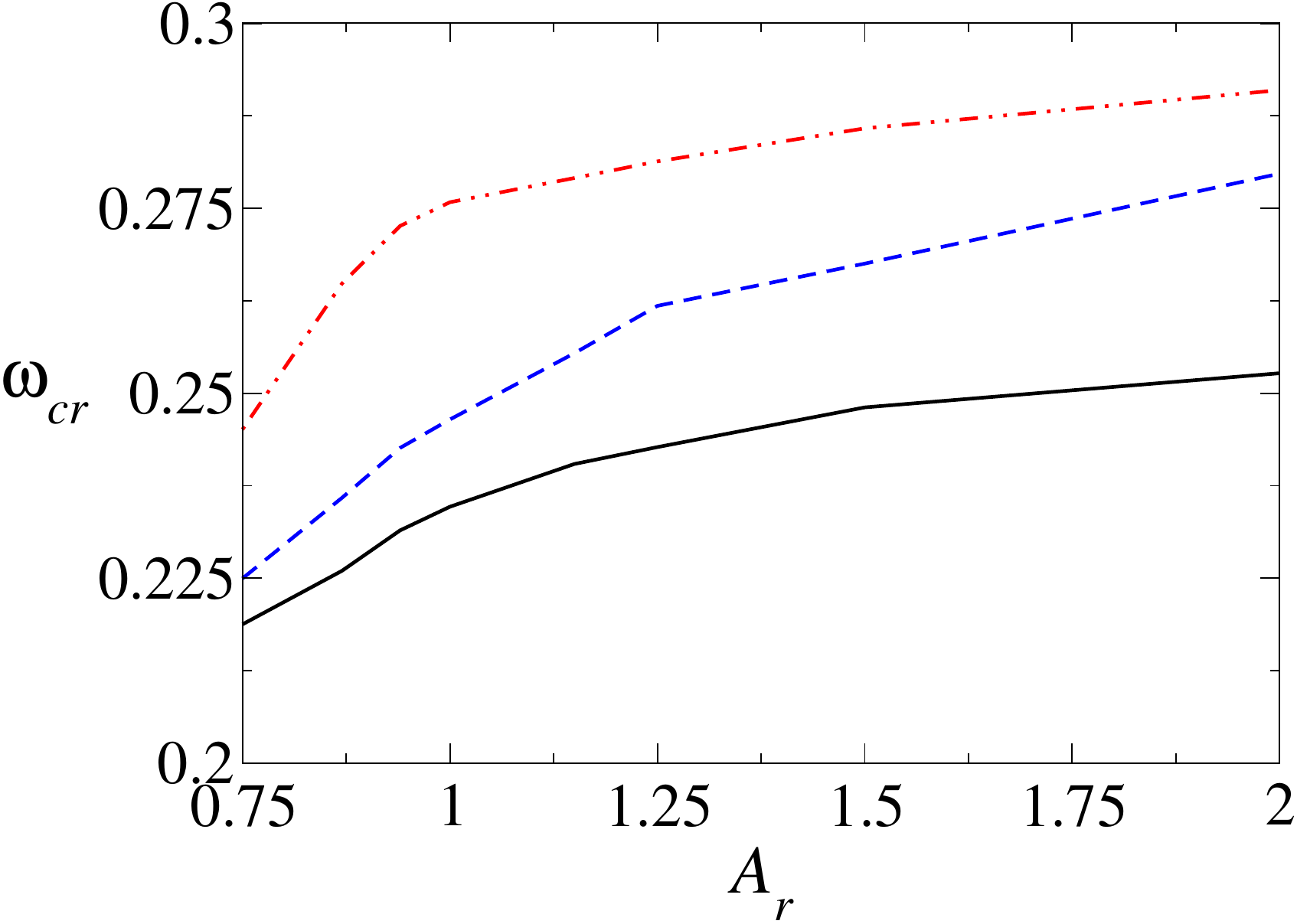}\hspace{0.5cm}
\includegraphics[scale=0.18]{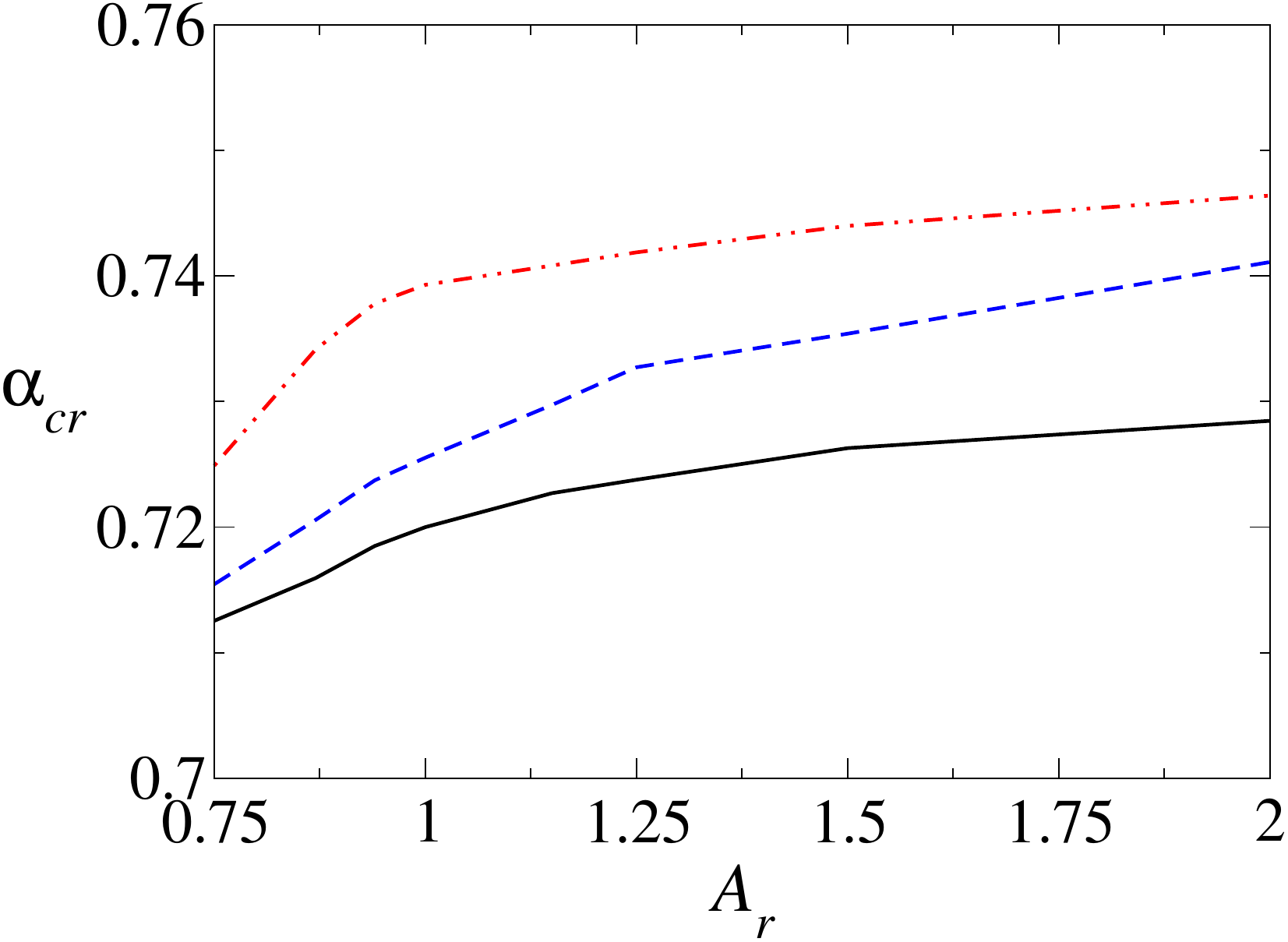}
\caption{(a) Variation of the critical growth rate $(\omega_{cr})$ and (b) critical wavenumber $(\alpha_{cr})$ with aspect ratio $(A_r)$ for $We = 729$ at different time instances $\tau = 42.9$, 47.6, and 52.4.}
\label{fig:analytical_v}
\end{figure}

There has been considerable debate regarding the primary mechanism responsible for node formation at the rim of a crown during splashing. Some studies attribute it to Rayleigh–Plateau (RP) instability \citep{zhang2010wavelength}, others to Rayleigh–Taylor (RT) instability \citep{krechetnikov2009crown}, while some propose a combination of both these instability mechanisms \citep{roisman2006spray}. In this section, we aim to identify the dominant instability mechanism governing node formation for the system considered in the present study. The Rayleigh–Plateau (RP) instability manifests itself as symmetric perturbations along the rim axis, driven by variations in the axial velocity ($u$) and the radius of the rim ($r$). In contrast, perturbations in $Q$ and $Y$ lead to a bending of the rim centerline \citep{roisman2006spray}. To isolate the Rayleigh–Plateau (RP) instability, we introduce perturbations $\hat{\eta}$ and $\hat{u}$, while setting $\hat{\epsilon} = 0$ and $\hat{q} = 0$. Under these assumptions, the dispersion relation from eq.~\eqref{dispersion_eqn} simplifies to:
\begin{equation}
2\omega^2 + W_0\omega + \alpha^2 (\alpha^2 - 1) = 0. \label{RP_dispersion}
\end{equation}
In contrast, for analyzing the Rayleigh–Taylor (RT) instability, we introduce perturbations $\hat{\epsilon}$ and $\hat{q}$, while keeping $\hat{\eta} = 0$ and $\hat{u} = 0$. This is because the RT instability develops perpendicular to the rim axis, in the direction of its deceleration, and is triggered solely by perturbations in $Y$ and $Q$ \citep{roisman2006spray}. Under these conditions, the dispersion relation derived from eq.~\eqref{dispersion_eqn} simplifies to:
\begin{equation}
\pi(2+\alpha^2)\omega^2 + W_0(4+\alpha^2)\omega + 2\pi\alpha^2 = 0. \label{RT_dispersion}
\end{equation}

\begin{figure}
\centering
\includegraphics[scale=0.3]{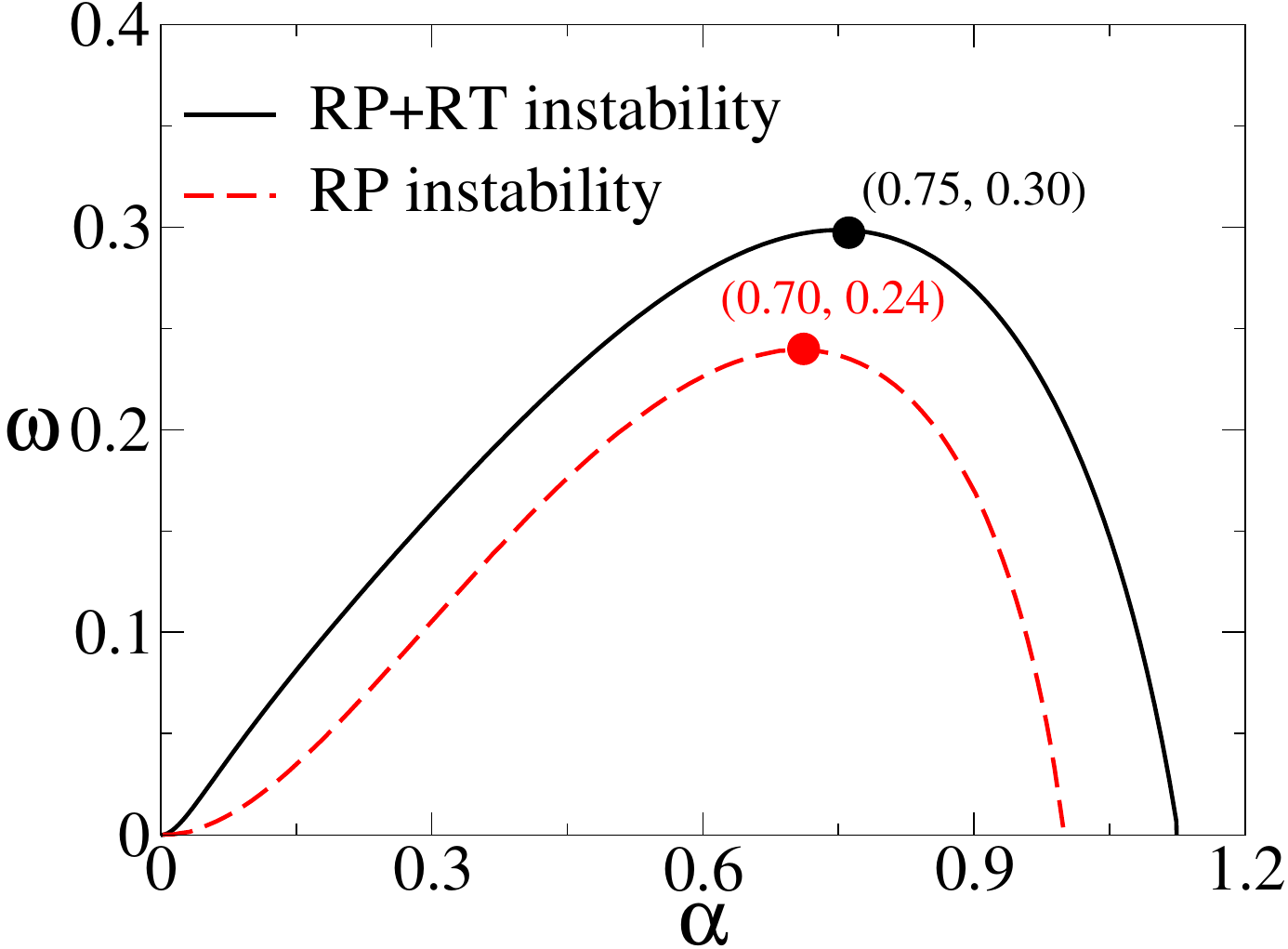}
\caption{Dispersion curves ($\omega$ versus $\alpha$) of the RP instability (dashed red line) and the combined RP+RT instability (solid black line) for $A_r=1.0$, $\We=1050$ at $\tau = 71.43$.}
\label{fig:RP_dispersion}
\end{figure}

The dispersion curve obtained from the simplified RP instability relation (eq.~\ref{RP_dispersion}) is shown in figure~\ref{fig:RP_dispersion}. It exhibits a parabolic shape, with $\omega > 0$ over a finite wavenumber range, indicating linear instability. The curve exhibits a distinct “most dangerous” mode, corresponding to the wavenumber ($\alpha_{cr}$) at which the growth rate is maximal ($\omega_{cr}$), and a “cutoff” mode, beyond which disturbances decay ($\omega < 0$). In contrast, the RT instability relation (eq.~\ref{RT_dispersion}) yields no positive real roots for $\omega$, implying that this mode is stable and does not independently lead to node formation. This observation is consistent with the findings of \citet{roisman2006spray}. Hence, the formation of nodes is primarily attributed to the RP instability. However, when both RP and RT instabilities are considered by solving the full dispersion relation (eq.~\ref{dispersion_eqn}), the resulting growth rate is higher than that from the RP mode alone (see figure~\ref{fig:RP_dispersion}). Notably, this trend is further supported by the data in Table~\ref{tab:RPvsRT_table}, which shows that the most unstable wavelength remains nearly unchanged between the two cases when compared to the change in the growth rate. This behavior is consistent across a range of aspect ratios ($A_r$), as presented in Table~\ref{tab:RPvsRT_table}, which provides a quantitative comparison of $\omega_{cr}$ and $\alpha_{cr}$ at different instants for both instability modes. These results suggest that RP instability determines the dominant wavelength and number of fingers, while RT instability significantly alters the growth rate of the perturbations. Thus, RP instability governs the spatial pattern of node formation, while RT instability contributes primarily to their amplification. In addition to the planar crown assumption used in the linear analysis presented in this section, we have also performed a curved-rim stability analysis to assess the effect of crown curvature on the dispersion characteristics. Our results show no significant change in either the maximum growth rate ($\omega_{cr}$) or the corresponding wavenumber ($\alpha_{cr}$). A detailed discussion of this analysis has been included in Appendix~\ref{sec:curved_rim}.

\begin{table}
\centering
\begin{tabular}{cccccccc}
\multirow{2}{*}{$A_r$} ~~~ & \multicolumn{3}{c}{$\omega_{cr}$} & \multicolumn{3}{c}{$\alpha_{cr}$} & ~~~ \multirow{2}{*}{ time ($\tau$)}\\
\cline{2-7}\\
 & $RP$ & $RP + RT$ & ${\cal A}$ & $RP$ & $RP + RT$ & ${\cal B}$ \\
\hline
\multirow{3}{*}{0.75} 
 & 0.234 & 0.238 & 2.01\% & 0.707 & 0.714 & 1.04\% & 38.1\\
 & 0.233 & 0.238 & 2.30\% & 0.707 & 0.711 & 0.62\% & 42.9\\
 & 0.233 & 0.238 & 2.29\% & 0.707 & 0.710 & 0.48\% & 56.0\\
\hline
\multirow{3}{*}{1.0}  
 & 0.240 & 0.288 & 19.92\% & 0.707 & 0.746 & 5.58\% & 38.1\\
 & 0.235 & 0.248 & 5.77\% & 0.707 & 0.722 & 2.18\% & 42.9\\
 & 0.234 & 0.243 & 3.63\% & 0.707 & 0.718 & 1.61\% & 56.0\\
\hline
\multirow{3}{*}{1.5}  
 & 0.235 & 0.25 & 6.54\% & 0.707 & 0.746 & 2.75\% & 38.1\\
 & 0.238 & 0.272 & 14.26\% & 0.707 & 0.734 & 3.88\% & 42.9\\
 & 0.243 & 0.306 & 25.85\% & 0.707 & 0.755 & 6.72\% & 56.0\\
\hline
\end{tabular}
\caption{Comparison of the maximum growth rate ($\omega_{cr}$) and the critical wavenumber ($\alpha_{cr}$) obtained from eq.~\eqref{RP_dispersion} (RP mode) and eq.~\eqref{dispersion_eqn} (combined RP and RT modes) for  Weber numbers ($\We = 729$) for various aspect ratios ($A_r$) at different time instants. Here, $D_{crown}$ denotes the crown diameter, ${\cal A} = \left (\omega_{cr, RP+RT}-\omega_{cr, RP})/\omega_{cr, RP} \right ) \times 100$ and ${\cal B} = \left (\alpha_{cr, RP+RT}-\alpha_{cr, RP})/\alpha_{cr, RP} \right ) \times 100$. The analysis is conducted for a water–air fluid system.}
\label{tab:RPvsRT_table}
\end{table}

\begin{figure}
\centering
\includegraphics[scale=0.7]{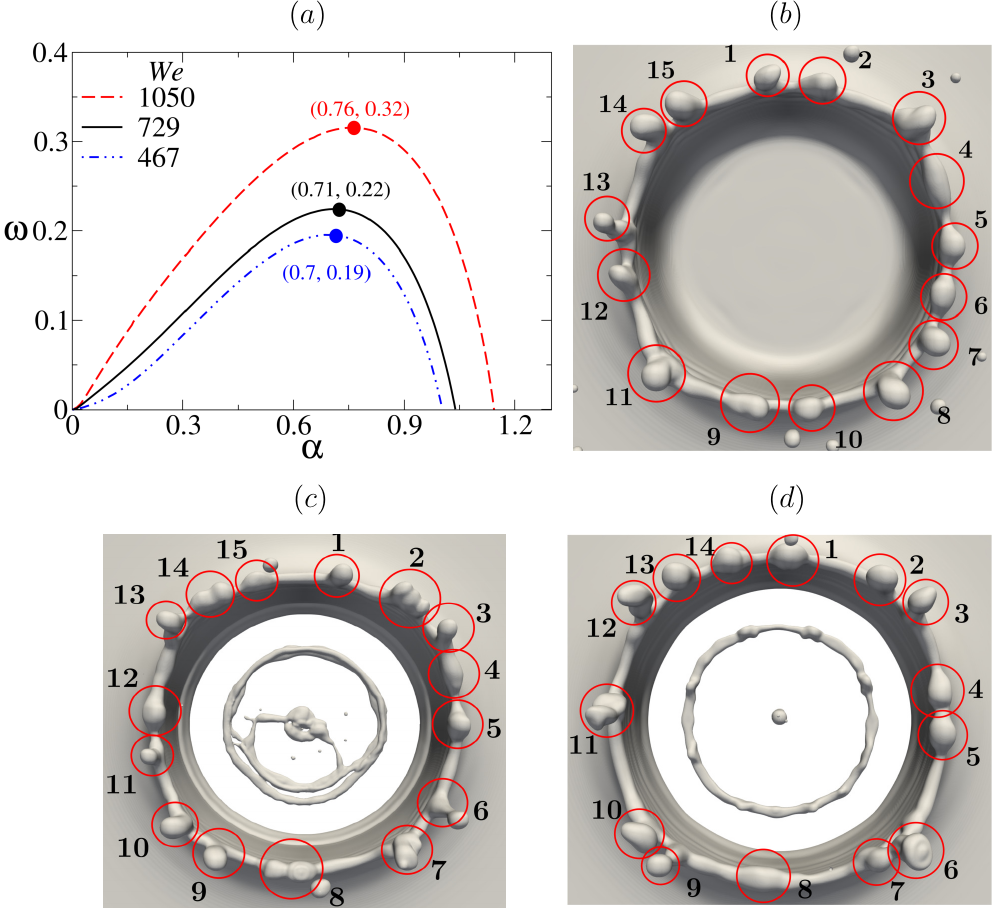}
\caption{$(a)$ Dispersion curves for $A_r = 2.0$ at $\We = 467$, 729, and 1050. The number of nodes corresponding to the most unstable mode for $(b)$ $\We = 467$ at $\tau = 25.7$, $(c)$ $\We = 729$ at $\tau = 36.9$, and $(d)$ $\We = 1050$ at $\tau = 57.1$. The identified nodes are highlighted in red and numbered accordingly. The fluid system considered here is a water-air system.}
\label{fig:LSA_Ar_2}
\end{figure}

\subsection{Parametric study}

Figure~\ref{fig:LSA_Ar_2}($a$) shows the dispersion curves ($\omega$ versus $\alpha$) for various Weber numbers ($\We$) obtained from linear stability analysis. The respective dispersion curves indicate the values of $(\alpha_{cr}, \omega_{cr})$ corresponding to each Weber number. It can be seen that increasing $\We$ increases the growth rate of the most unstable mode ($\omega_{cr}$), indicating that higher impact velocities intensify the instability of the crown rim. A higher $\omega_{cr}$ implies that small perturbations on the rim grow more rapidly, thereby increasing the likelihood of node formation. This observation is consistent with the regime boundaries shown in figure~\ref{fig:Regime_map}, where higher $\We$ values correspond to more frequent splashing events. These unstable nodes can evolve into fingers, eventually detaching as secondary droplets, which is a distinctive feature of the splashing process. Figure~\ref{fig:LSA_Ar_2}($b–d$) show the snapshots obtained from the numerical simulations highlighting the number of nodes for $\We = 467$, 729, and 1050, respectively. The nodes are circled in red and numbered. The number of nodes observed for different Weber numbers agrees well with the theoretical prediction from the linear stability analysis. Interestingly, we observe that the number of nodes does not increase monotonically with $\We$, despite the higher growth rates predicted by the linear stability analysis. This suggests that while a higher $\We$ increases the growth rate of the most unstable mode (thereby accelerating crown development), the node formation is affected by other factors, such as the rim radius ($r_0$) and local crown irregularities, including hole ruptures, which can disrupt regular node formation. 

\begin{figure}
\centering
\hspace{0.5cm} $(a)$ \hspace{5.8cm} $(b)$ \\
\includegraphics[width=0.45 \textwidth]{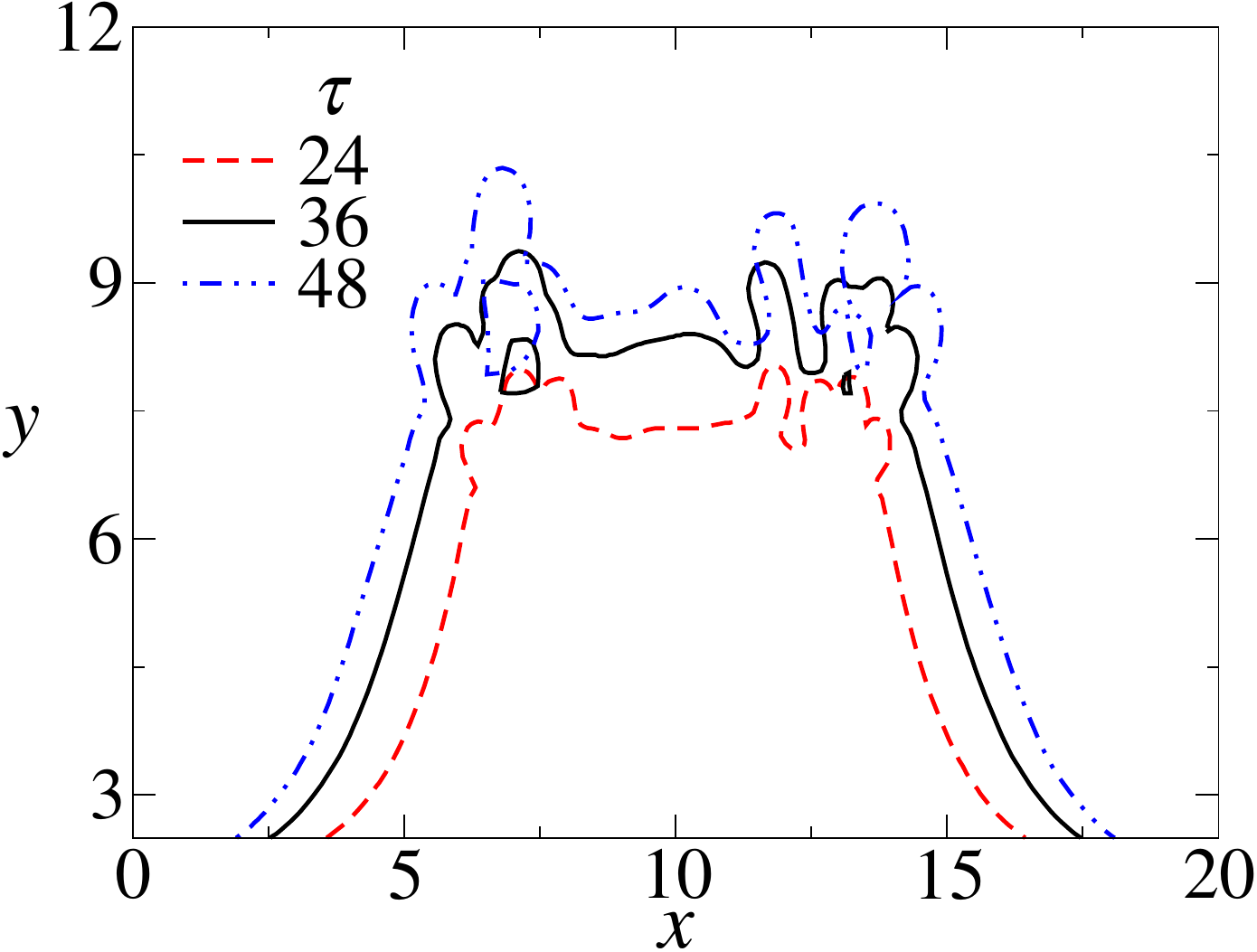} \hspace{2mm}
\includegraphics[width= 0.45\textwidth]{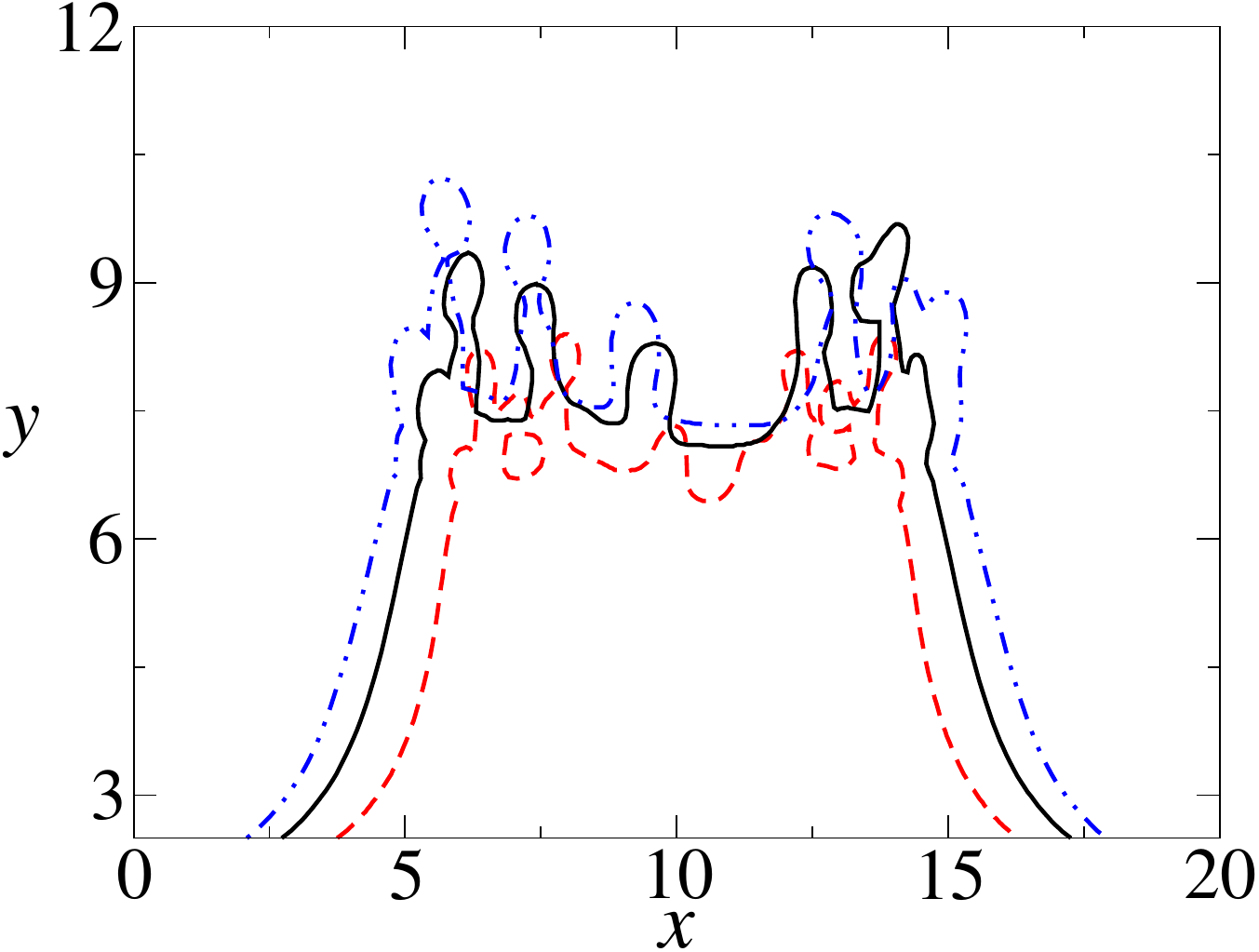} \\
\hspace{0.5cm} $(c)$ \hspace{5.8cm} $(d)$ \\
\includegraphics[width= 0.45 \textwidth]{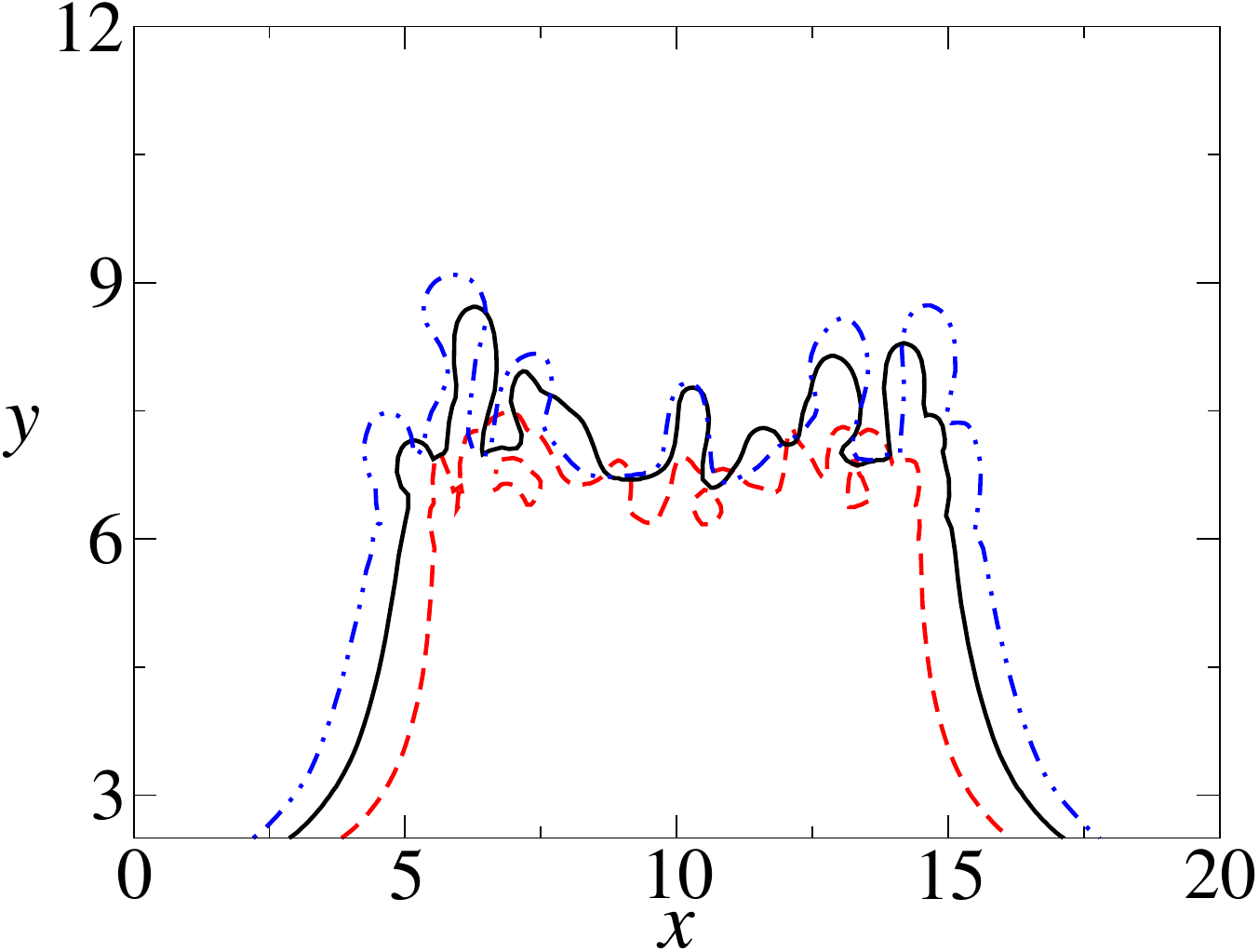} \hspace{2mm}
\includegraphics[width=0.45 \textwidth]{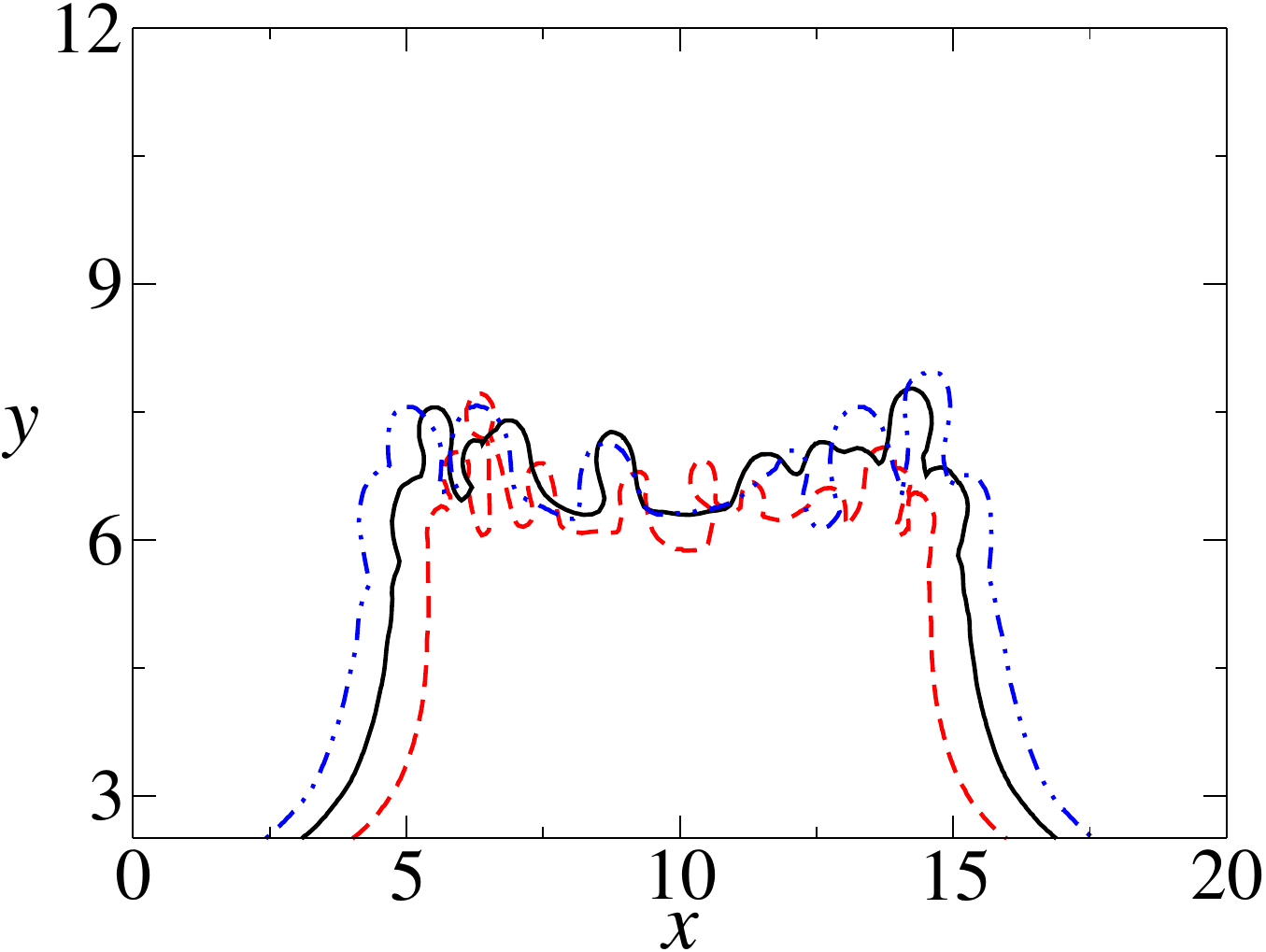}
\caption{Evolution of the front view of the crown for $(a)$ $A_r = 0.75$, $(b)$ $A_r = 1.0$, $(c)$ $A_r = 1.5$, and $(d)$ $A_r = 2.0$, for $\We = 729$ at $\tau =  24$, 36, 48. Here, $x$ and $y$ are dimensionless horizontal and vertical distances, respectively. The fluid system considered here is a water-air system.} 
\label{fig:crown_evolution}
\end{figure}

Building on the insights from figure~\ref{fig:LSA_Ar_2}, which illustrated the relationship between the growth rate ($\omega$), Weber number ($\We$), and the number of nodes ($n$) along the crown rim, figure~\ref{fig:crown_evolution} presents the temporal evolution of crown morphology for different aspect ratios ($A_r$) at $\We = 729$. This figure further highlights the influence of drop shape on crown dynamics by showing front-view snapshots of the crown at various non-dimensional times. Figure \ref{fig:crown_evolution}$(a-d)$ illustrate how the crown shape evolves for aspect ratios $A_r = 0.75$, 1.0, 1.5, and 2.0 respectively. At early times ($\tau = 24$, red dashed line), the crown forms a relatively smooth structure with small perturbations. As the time progresses to $\tau = 36$ (black solid line) and further to $\tau = 48$ (blue dash-dotted line), the crown develops more pronounced undulations. Additionally, the temporal difference in height reached by the rim of the crown is significantly higher for prolate drops compared to oblate drops. This suggests that oblate drops experience higher rim deceleration ($\dot{V_0}$) over a longer duration compared to prolate drops. This amplifies the RT instability, increasing the growth rate of nodes and promoting subsequent finger formation. This trend is consistent with the observation in figure~\ref{fig:Regime_map}, which shows that an increase in $A_r$ makes the drop more prone to splashing.

\begin{figure}
\centering
\hspace{0.3cm}$(a)$ \hspace{3.5cm} $(b)$ \hspace{3.5cm} $(c)$\\
\includegraphics[scale=0.18]{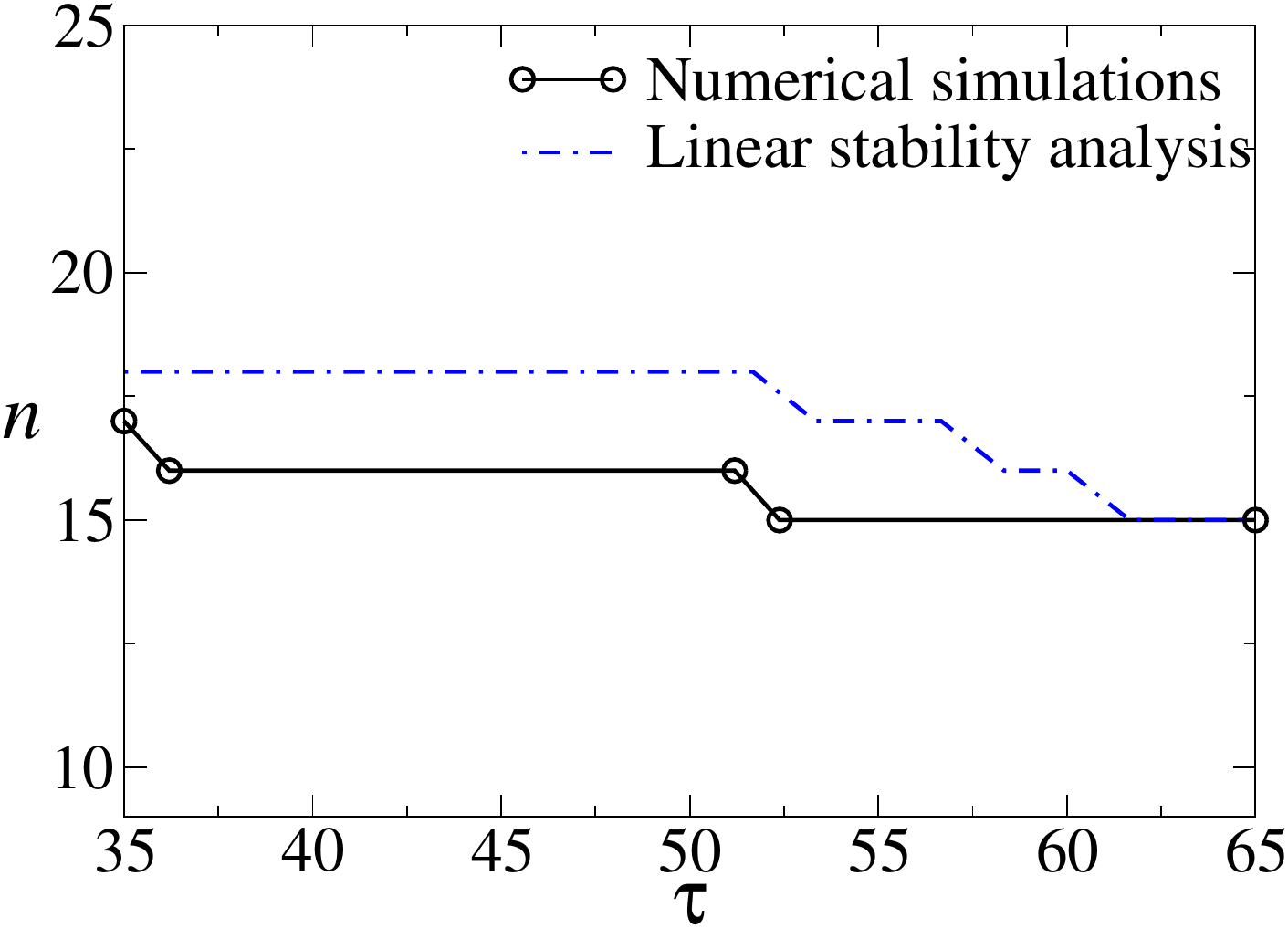}
\includegraphics[scale=0.18]{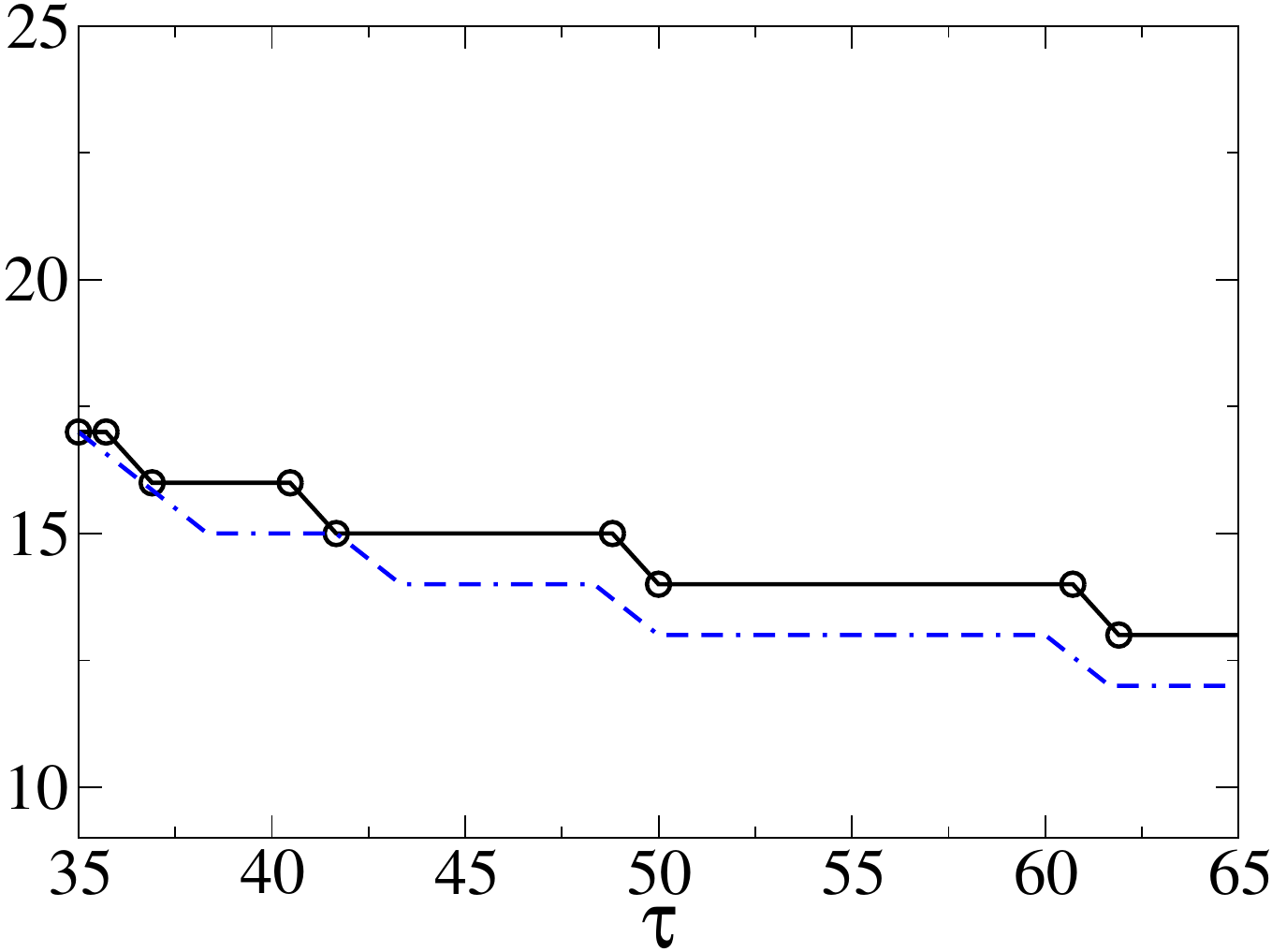}
\includegraphics[scale=0.18]{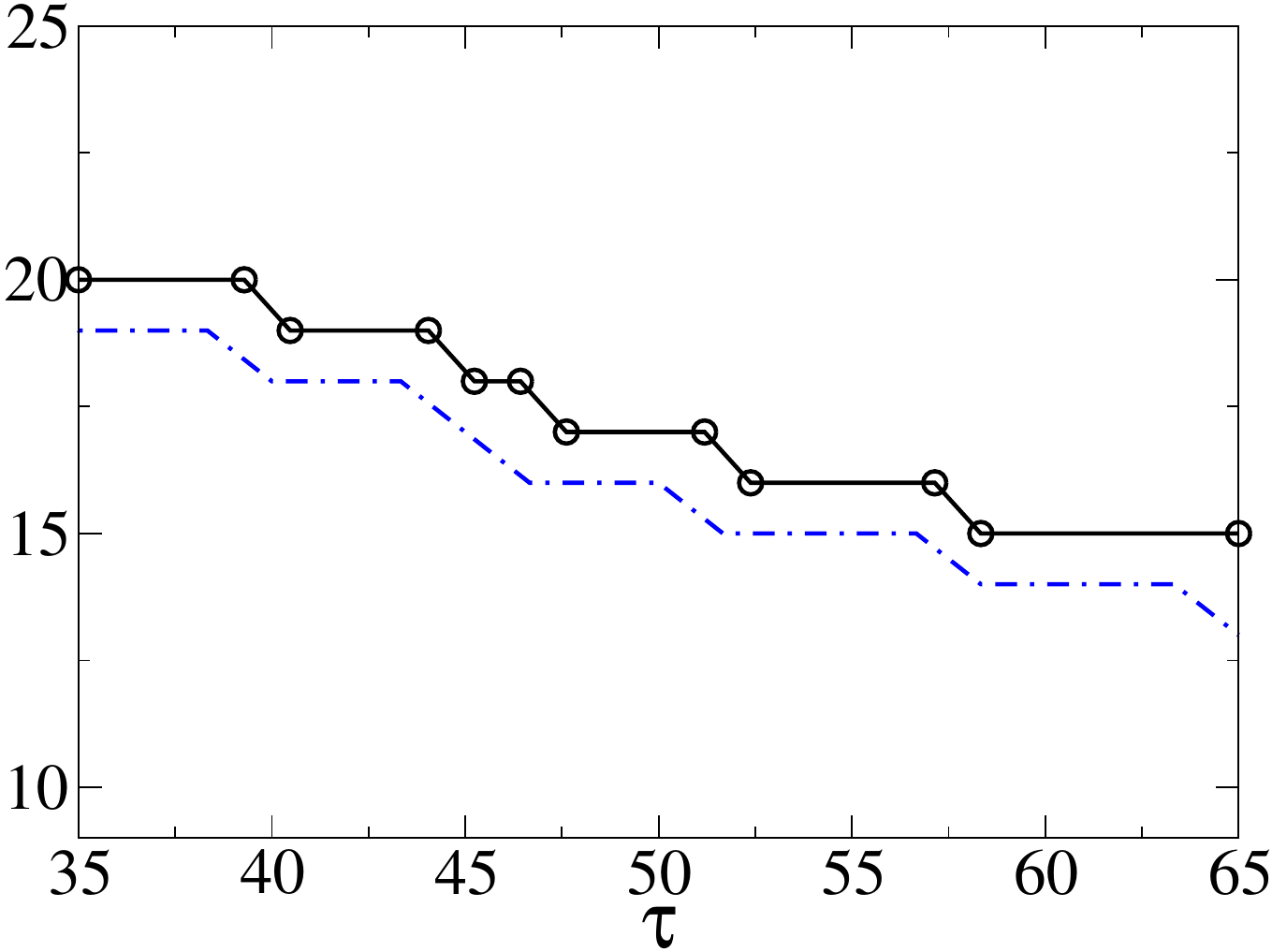}
\caption{Comparison of the predicted number of nodes from the linear stability analysis (blue dash–dot line) with the observed number of nodes from numerical simulations (black solid line) for $\We = 729$ at $(a)$ $A_r = 0.75$, $(b)$ $A_r = 1.0$, and $(c)$ $A_r = 1.5$.}
\label{fig:nodes_comp}
\end{figure}

It is important to note that while linear stability analysis provides a crucial estimate for the onset of splashing by predicting the number of nodes formed on the crown rim using the critical wave number ($\alpha_{cr}$), not all of these nodes evolve into fully developed fingers. Nonlinear effects, such as merging and splitting of nodes induced by holes in the crown (figure~\ref{fig:Hole_demonstration}), play a crucial role in governing the subsequent evolution of the rim structure. Additionally, as shown in figure~\ref{fig:crown_evolution}, while some of the initially formed nodes evolve into nodes, others diminish over time and do not develop into visible fingers. This damping effect results from the complex interplay between the local momentum distribution along the rim and the instability growth rate. Specifically, insufficient fluid accumulation in certain rim regions suppresses node growth, preventing them from evolving into fingers.

Figure~\ref{fig:nodes_comp} illustrates the temporal evolution of the nodes predicted by linear stability analysis for droplets with aspect ratios $A_r = 0.75$, 1.0, and 1.5 at $\We=729$. It is observed that the number of nodes is not constant but decreases over time. This trend is clearly visible in the figure~\ref{fig:nodes_comp}, where a steady decline in node count can be observed, even as the crown diameter $D_{crown}$ continues to grow. The observed reduction in node count is attributed to the accumulation of fluid from the lamella into the rim, which increases the rim radius $(r_0)$ over time, as also seen in figure \ref{fig:extracted_v}$b$. As $r_0$ increases, the most unstable wavelength grows longer, resulting in fewer nodes forming around the rim. Notably, the number of nodes predicted by linear theory represents an upper bound, with not all predicted nodes necessarily evolving into visible fingers. The conversion of nodes into fingers depends on rim deceleration, which governs the growth rate of the Rayleigh–Taylor (RT) instability. Additionally, the local redistribution of fluid within the rim further modulates this transformation, underscoring the complex interaction between rim geometry and the growth of instability. For the prolate case ($A_r = 0.75$), the comparison in figure~\ref{fig:nodes_comp} shows that the predicted node count remains higher than the value observed in the simulations. This behaviour differs slightly from the spherical and oblate cases, where the predicted count is lower than the simulated value. This can be explained by the combined effect of the rim radius and crown diameter. Since the prolate drop has the smallest rim radius ($r_0$), it exhibits a higher critical wavenumber ($\alpha_{cr}$), leading to a slight overprediction of the number of nodes in the linear stability analysis. At the same time, the crown diameter ($D_{crown}$) is also the smallest, which leads to closer spacing between neighbouring fingers. As the crown evolves, these closely spaced fingers merge more readily, resulting in a smaller number of distinct nodes in the simulation. Also in prolate drops, weaker deceleration results in reduced fluid accumulation at the rim and a minor increase in $r_0$ over time (as observed in figure \ref{fig:extracted_v}$b$). Consequently, the most unstable wavelength remains nearly constant, and the number of predicted nodes exhibits minimal temporal variation. In contrast, oblate droplets experience more substantial deceleration, which drives more fluid into the rim, significantly increasing $r_0$ and reducing the number of nodes over time. This behavior is consistently captured in both theoretical predictions and simulations. Initially, the oblate drop, despite having a higher rim radius compared to the prolate drop, has a higher number of nodes due to its increased rim circumference. However, due to the rapid increase in rim radius driven by high deceleration, the node count in the oblate case eventually decays much faster than that of the prolate and spherical drops.

\section{Concluding remarks} \label{sec:conc}
We have performed a comprehensive three-dimensional numerical study of the splashing dynamics of non-spherical droplets impacting a quiescent liquid film, covering a broad range of aspect ratios ($A_r$) and Weber numbers ($\We$). Our numerical simulations identify distinct impact phenomena, such as spreading, splashing type-1, splashing type-2, and canopy formation, and present a regime map demarcating these impact outcomes in the $A_r$–$\We$ parameter space. This regime map highlights the influence of shape-induced anisotropy and inertial effects on transitions between different dynamical behaviors. Our results show that non-spherical droplets produce significantly different crown structures and splashing patterns compared to spherical droplets. Specifically, oblate droplets experience greater rim deceleration, making them more susceptible to fingering and fragmentation, while prolate droplets tend to delay the onset of instability and trigger canopy formation due to their narrower crown bases and enhanced vertical expansion. Furthermore, the splashing threshold, characterized by a critical Weber number, decreases with increasing $A_r$, underscoring the sensitivity of splash dynamics to droplet shape and impact conditions. A key aspect of this work is the integration of numerical simulations with linear stability analysis to predict the number of nodes that evolve into fingers along the crown rim. By directly extracting physical parameters, such as rim radius and acceleration, from the numerical simulations, the stability analysis adequately captures the influence of both Rayleigh–Plateau (RP) and Rayleigh–Taylor (RT) instabilities on finger development. Our findings demonstrate that the RP mechanism governs the characteristic wavelength and number of undulations, while the RT mechanism enhances their growth rate. The theoretical analysis combining RP and RT modes agrees well with numerical observations. Furthermore, the present study reveals the time-dependent nature of finger development, with the number of fingers varying throughout the splashing process due to nonlinear effects such as hole formation, rim breakup, and fluid redistribution. 

\appendix
\section{Linear stability analysis for curved rim}\label{sec:curved_rim}

\begin{figure}
\centering
\hspace{1cm} 
\includegraphics[scale=0.6]{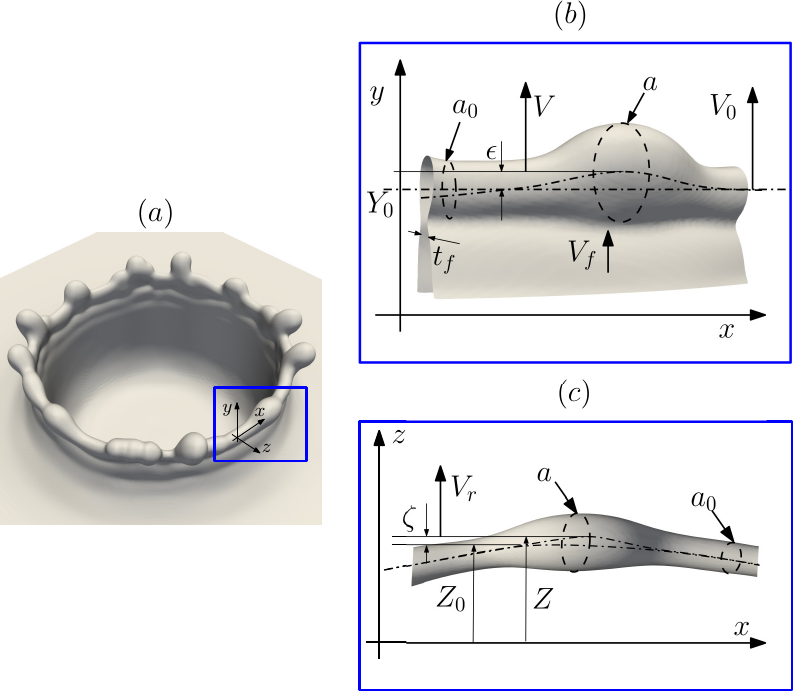}
\caption{(a) A representative crown structure obtained from numerical simulation. (b) The perturbed state of the rim (base state plus disturbance) in the $z$-direction. The subscript ``0'' denotes quantities corresponding to the unperturbed base state, whereas perturbed quantities employed in the linear stability analysis are indicated without subscripts.}
\label{fig:curved_LSA_sche}
\end{figure}

\quad In addition to the planar crown assumption used in the linear stability analysis, a curved-rim analysis was performed to examine the influence of crown curvature on the number of fingers. The rim is modeled as a toroidal liquid element of radius $r_0$, parameterized along the $x$-direction (axis of the rim), with the lamella thickness $t_f$ feeding the rim. By incorporating the radial curvature of the rim, the quantities in the $z$-direction (see Figure~\ref{fig:curved_LSA_sche}) are also considered. The corresponding governing equations are given by
\begin{subequations}
\begin{equation}
\frac{\partial a}{\partial t} + {a_0} \frac{\partial u}{\partial x} - {t_f}( V_f - V) = 0, \label{dim_mass_con}
\end{equation}
\begin{equation}
\rho {a_0} \frac{\partial u}{\partial t} - \frac{\partial P}{\partial x} - \left[ - \rho {t_f} (V_f -V) u + 2 \sigma \frac{\partial Y}{\partial x} +2\sigma\frac{\partial Z}{\partial x}\right ] = 0,
\label{dim_xmom_con}
\end{equation}
\begin{equation}
\rho a \frac{\partial V}{\partial t} - P \kappa - \frac{\partial Q}{\partial x} - \left[ \rho {t_f} (V_f - V)^2 - 2\sigma \right] = 0,
\label{dim_ymom_con}
\end{equation}
\begin{equation}
\rho \frac{\partial L}{\partial t} - \frac{\partial M}{\partial x} - Q - {m_S} = 0, \label{dim_xyamom_con}
\end{equation}
\begin{equation}
\rho a \frac{\partial V_r}{\partial t} - P \kappa_r - \frac{\partial Q_r}{\partial x} - 2\sigma = 0,
\label{dim_zmom_con}
\end{equation}
\begin{equation}
\rho \frac{\partial L_r}{\partial t} - \frac{\partial M_r}{\partial x} - Q_r = 0. \label{dim_xzamom_con}
\end{equation}
\end{subequations}
\quad Here, $\kappa$ and $\kappa_r$ denote the curvature of the rim centerline in the $z$- and $y$-axes, respectively; $Q$ and $Q_r$ represent the shear forces in the $y$- and $z$-directions acting on the rim cross-section; and $m_S$ is the distributed moment arising from external forces (primarily the flow from the lamella into the rim). The longitudinal tensile force ($P$), which includes both the capillary pressure and surface tension effects, is given by
\begin{equation}
P = \pi \sigma r + \sigma {a_0} \frac{\partial^2 r}{\partial x^2}.
\end{equation}
\quad The angular momenta of the rim per unit length ($L$ and $L_r$) are
\begin{equation}
L = I_0 \Omega, \quad L_r = I_0 \Omega_r, \quad \text{where} \quad I_0 = \frac{\pi {r_0}^4}{4}.
\end{equation}
The internal moments ($M$, $M_r$) resulting from stresses due to pressure gradients in the $y$- and $z$-directions are
\begin{equation}
M = -\rho I_0 \frac{d V}{d t}, \quad M_r = -\rho I_0 \frac{d V_r}{d t}.
\end{equation}
The moment due to lamella inflow is
\begin{equation}
m_S = -\rho {t_f} {r_0} (V_f - V_0) u.
\end{equation}

\begin{figure}
\centering
\includegraphics[scale=0.3]{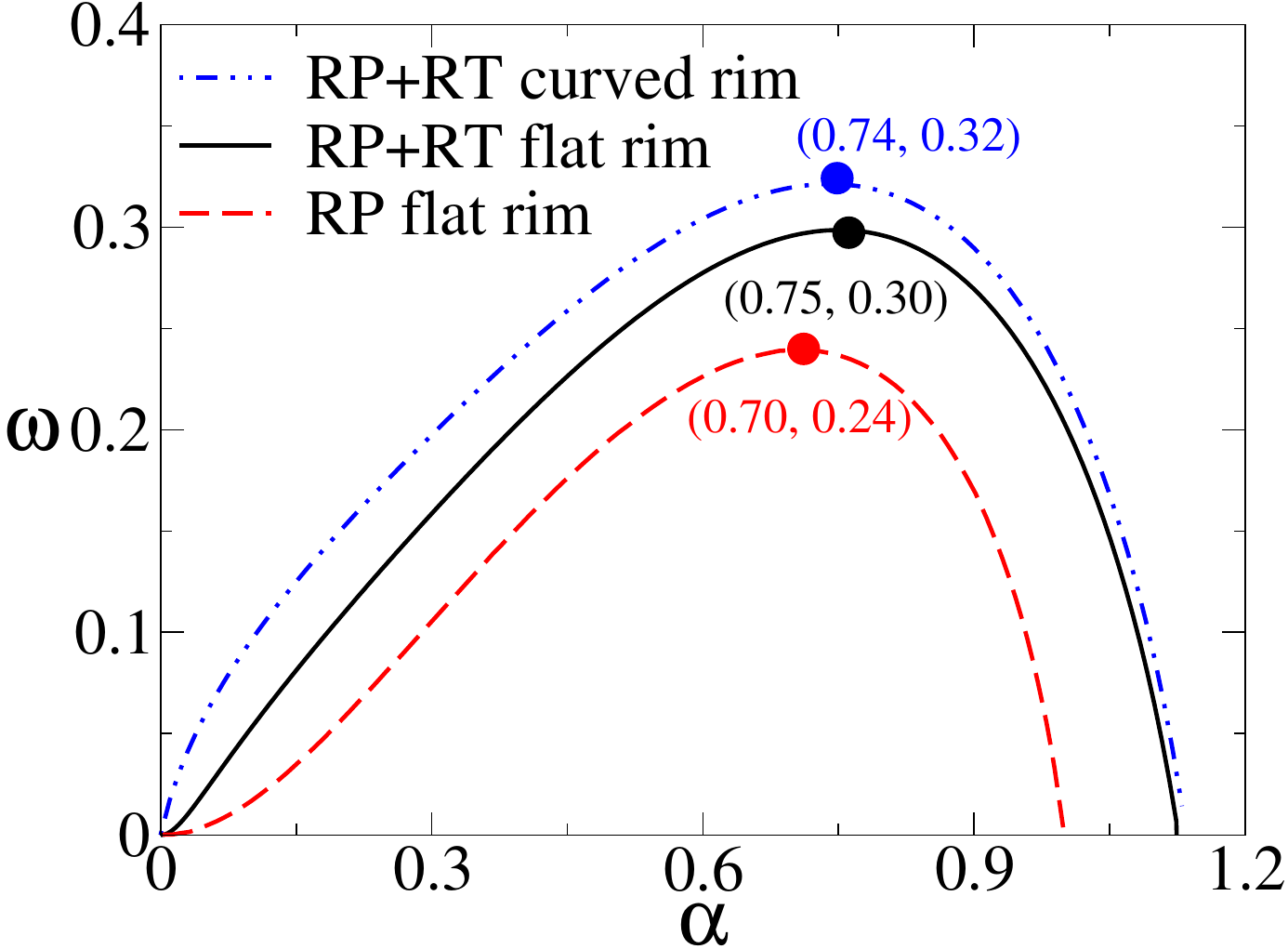}
\caption{Dispersion curves ($\omega$ versus $\alpha$) for the combined RP+RT instability with the curved-rim assumption (dash-dotted blue line), the combined RP+RT instability with the flat-rim assumption (solid black line), and the RP instability alone (dashed red line) at $A_r = 1.0$, $We = 1050$, and $\tau = 71.43$.}
\label{fig:curved_dispersion}
\end{figure}

Following the standard linear stability framework, each variable is represented as the sum of a base state and a small perturbation decomposed into its amplitude and wave components:
\begin{equation}
(Y,r,u,Q,Z,Q_r)(x,y,t) = [Y_0, r_0, 0, 0, Z_0, 0] + \left(\hat{\epsilon}, \hat{\eta}, \hat{u}, \hat{q}, \hat{\zeta}, \hat{q_r}\right) \exp(\omega_c t + i \alpha x),
\label{perturbations_curved}
\end{equation}
where the hats denote perturbation amplitudes, $\alpha = 2\pi / \lambda$ is the real wavenumber, and $\omega_c$ is the complex frequency. The real part of $\omega_c$, denoted by $\omega = \text{Re}(\omega_c)$, represents the growth rate; thus, $\omega > 0$ corresponds to an unstable mode, $\omega < 0$ to a stable mode, and $\omega = 0$ to neutral stability. Assuming slow temporal variations of $r_0$, $a_0$, and $I_0$, a quasi-steady approximation is adopted. Substituting eq.~(\ref{perturbations_curved}) into the governing equations and linearizing them yield the dispersion relation:
\begin{eqnarray}
&&
\Big(
 -4 \pi^3 {d {\bar V_0} \over dt} \bar\alpha^4 
 - 4 \pi^3 {d {\bar V_{r0}} \over dt} \bar\alpha^4
 + \pi^4 \bar\alpha^6
 - \pi^3 {d {\bar V_0} \over dt} \bar\alpha^6
 - \pi^3 V_{r0} \bar\alpha^6
 - \pi^4 \bar\alpha^8
\Big)
\notag \\
&&+
\Big(
 2 \pi^2 {\bar t_f} {d {\bar V_0} \over dt} {\bar W}_0 \bar\alpha^2
 - 8 \pi^2 {d {\bar V_{r0}} \over dt} {\bar W}_0 \bar\alpha^2
 - 4 \pi {\bar t_f} {d {\bar V_{r0}} \over dt} {\bar W}_0 \bar\alpha^2
 - 4 \pi^2 {\bar W}_0 \bar\alpha^4
 + \pi^2 {\bar t_f} {\bar W}_0 \bar\alpha^4
\notag \\
&&\qquad
 + 2 \pi^2 {d {\bar V_0} \over dt} {\bar W}_0 \bar\alpha^4
 + 0.5 \pi^2 {\bar t_f} {d {\bar V_0} \over dt} {\bar W}_0 \bar\alpha^4
 - 4 \pi^2 {d {\bar V_{r0}} \over dt} {\bar W}_0 \bar\alpha^4
 - \pi {\bar t_f} {d {\bar V_{r0}} \over dt} {\bar W}_0 \bar\alpha^4
\notag \\
&&\qquad
 - 1.5 \pi^3 {\bar W}_0 \bar\alpha^6
 - \pi^2 {\bar t_f} {\bar W}_0 \bar\alpha^6
 + 0.5 \pi^2 {d {\bar V_0} \over dt} {\bar W}_0 \bar\alpha^6
 - 0.5 \pi^2 {d {\bar V_{r0}} \over dt} {\bar W}_0 \bar\alpha^6
 - 0.5 \pi^3 {\bar W}_0 \bar\alpha^8
\Big)\bar\omega
\notag \\
&&+
\Big(
 -4 \pi^3 {d {\bar V_0} \over dt} \bar\alpha^2
 + 2 \pi^3 {\bar t_f} {d {\bar V_0} \over dt} \bar\alpha^2
 - 4 \pi^3 {d {\bar V_{r0}} \over dt} \bar\alpha^2
 - 4 \pi^2 [{\bar t_f}(2+\pi {d{\bar V_0}\over dt})]\bar\alpha^2
\notag \\
&&\qquad
 - 3 \pi^3 {d{\bar V_0}\over dt}\bar\alpha^4
 + 0.5\pi^3{\bar t_f}{d{\bar V_0}\over dt}\bar\alpha^4
 - 3\pi^3{d{\bar V_{r0}}\over dt}\bar\alpha^4
 + 2\pi[{\bar t_f}(2+\pi{d{\bar V_0}\over dt})]\bar\alpha^4
\notag \\
&&\qquad
 - \pi^2 [{\bar t_f}(2+\pi{d{\bar V_0}\over dt})]\bar\alpha^4
 - \pi^4\bar\alpha^6
 - 0.5\pi^3{d{\bar V_0}\over dt}\bar\alpha^6
 - 0.5\pi^3{d{\bar V_{r0}}\over dt}\bar\alpha^6
 - \pi^4\bar\alpha^8
\Big)\bar\omega^2
\notag \\
&&+
\Big(
 2\pi^2{\bar t_f}{d{\bar V_0}\over dt}{\bar W}_0
 - 4\pi^2{\bar W}_0\bar\alpha^2
 - 6\pi^3{\bar W}_0\bar\alpha^2
 + \pi^2{\bar t_f}{\bar W}_0\bar\alpha^2
\notag \\
&&\qquad
 + 1.5\pi^2{\bar t_f}{d{\bar V_0}\over dt}{\bar W}_0\bar\alpha^2
 - 2\pi^2{\bar W}_0\bar\alpha^4
 - 3.5\pi^3{\bar W}_0\bar\alpha^4
 - 0.5\pi^2{\bar t_f}{\bar W}_0\bar\alpha^4
\notag \\
&&\qquad
 + 0.25\pi^2{\bar t_f}{d{\bar V_0}\over dt}{\bar W}_0\bar\alpha^4
 - 1.25\pi^3{\bar W}_0\bar\alpha^6
 - 0.5\pi^2{\bar t_f}{\bar W}_0\bar\alpha^6
 - 0.25\pi^3{\bar W}_0\bar\alpha^8
\Big)\bar\omega^3
\notag \\
&&+
\Big(
 2\pi^3{\bar t_f}{d{\bar V_0}\over dt}
 - 4\pi^2[{\bar t_f}(2+\pi{d{\bar V_0}\over dt})]
 - 3\pi^4\bar\alpha^2
 + 1.5\pi^3{\bar t_f}{d{\bar V_0}\over dt}\bar\alpha^2
\notag \\
&&\qquad
 - 3\pi^2[{\bar t_f}(2+\pi{d{\bar V_0}\over dt})]\bar\alpha^2
 - 2\pi^4\bar\alpha^4
 + 0.25\pi^3{\bar t_f}{d{\bar V_0}\over dt}\bar\alpha^4
 - 0.5\pi^2[{\bar t_f}(2+\pi{d{\bar V_0}\over dt})]\bar\alpha^4
\notag \\
&&\qquad
 - 0.75\pi^4\bar\alpha^6
 - 0.25\pi^4\bar\alpha^8
\Big)\bar\omega^4
\notag \\
&&+
\Big(
 -6\pi^3{\bar W}_0
 - 5\pi^3{\bar W}_0\bar\alpha^2
 - \pi^3{\bar W}_0\bar\alpha^4
\Big)\bar\omega^5
\notag \\
&&+
\Big(
 -2\pi^4
 - 2\pi^4\bar\alpha^2
 - 0.5\pi^4\bar\alpha^4
\Big)\bar\omega^6
=0. \label{curved_dispersion}
\end{eqnarray}
\quad Here, the dimensionless quantities are defined as
\begin{equation}
\omega =  \bar{\omega} \sqrt{\frac{\sigma}{\rho r_0^3}}, ~ \alpha = \frac{\bar{\alpha}}{r_0}, ~ t_f = \bar{t_f} r_0, ~
\dot{V_0} = \frac{\sigma}{\rho r_0^2} \bar{\dot{V_0}}, ~
\dot{V_{r0}} = \frac{\sigma}{\rho r_0^2} \bar{\dot{V_{r0}}}, ~
\bar{W_0} = \frac{\rho t_f (V_f - V_0)}{\sqrt{\rho \sigma a_0}}.
\end{equation}
\\
\quad Simplifying eq.~(\ref{curved_dispersion}) gives a sixth-order polynomial in $\omega$:
\begin{equation}
c_0 + c_1 \omega + c_2 \omega^2 + c_3 \omega^3 + c_4 \omega^4 + c_5 \omega^5 + c_6 \omega^6 = 0,
\end{equation}
where the coefficients $c_0-c_6$, depend on $\alpha$, $\dot{V_0}$, $\dot{V_{r0}}$, and $t_f$. These coefficients are given by
\begin{align*}
c_0 &= \pi^4 (\alpha^6 - \alpha^8) + \pi^3 \dot{V_0} (-4 \alpha^4 - \alpha^6) + \pi^3 \dot{V_{r0}}(-4 \alpha^4 - \alpha^6), \\
c_1 &= \sqrt{t_f \cdot (2 + \pi \cdot \dot{V_0})} \left[ \pi^2 \alpha^2 (2 t_f \dot{V_0} - 8 \dot{V_{r0}}- 4 \pi t_f \dot{V_{r0}}) \right. \\
&\quad \left.+ \pi^2 \alpha^4 (-4 + t_f + 2 \dot{V_0} + 0.5 t_f \dot{V_0} - 4 \dot{V_{r0}}- \pi t_f \dot{V_{r0}}) \right. \\
&\quad \left. + \alpha^6 (-1.5 \pi^3 - \pi^2 t_f + 0.5 \pi^2 \dot{V_0} - 0.5 \pi^2 \dot{V_{r0}}) - 0.5 \pi^3 \alpha^8 \right], \\
c_2 &= \pi^3 \dot{V_0} \alpha^2 (-4 + 2 t_f) + \pi^3 \dot{V_{r0}}(-4 \alpha^2 - 3 \alpha^4 - 0.5 \alpha^6) + [t_f \cdot (2 + \pi \cdot \dot{V_0})] \alpha^2 (-4 \pi^2) \\
&\quad + [t_f \cdot (2 + \pi \cdot \dot{V_0})] \alpha^4 (2 \pi - \pi^2) - \pi^4 (\alpha^6 + \alpha^8), \\
c_3 &= \sqrt{t_f \cdot (2 + \pi \cdot \dot{V_0})} \left[ 2 \pi^2 t_f \dot{V_0} + \pi^2 \alpha^2 (-4 - 6 \pi + t_f + 1.5 t_f \dot{V_0}) \right. \\
&\quad \left. + \pi^2 \alpha^4 (-2 - 3.5 \pi - 0.5 t_f + 0.25 t_f \dot{V_0}) - \pi^3 \alpha^6 \left(1.25 + 0.5 \pi^{-1} t_f\right) - 0.25 \pi^3 \alpha^8 \right], \\
c_4 &= 2 \pi^3 t_f \dot{V_0} - 4 \pi^2 [t_f \cdot (2 + \pi \cdot \dot{V_0})] + \pi^2 \alpha^2 \left(-3 \pi^2 + 1.5 \pi t_f \dot{V_0} - 3 [t_f \cdot (2 + \pi \cdot \dot{V_0})]\right) \\
&\quad + \pi^2 \alpha^4 \left(-2 \pi^2 + 0.25 \pi t_f \dot{V_0} - 0.5 [t_f \cdot (2 + \pi \cdot \dot{V_0})]\right) - \pi^4 \alpha^6 (0.75 + 0.25 \alpha^2), \\
c_5 &= \pi^3 \sqrt{t_f \cdot (2 + \pi \cdot \dot{V_0})} (-6 - 5 \alpha^2 - \alpha^4), \\
c_6 &= \pi^4 (-2 - 2 \alpha^2 - 0.5 \alpha^4).
\end{align*}
\quad Substituting these expressions into eq.~(\ref{curved_dispersion}) yields the dispersion curves. As illustrated in figure \ref{fig:curved_dispersion}, the differences in the maximum growth rate ($\omega_{cr}$) and the corresponding critical wavenumber ($\alpha_{cr}$) between eq.~(3.12) of the revised manuscript and eq.~(\ref{curved_dispersion}) presented here are minimal, indicating that the rim curvature has a negligible influence on the number of fingers formed.

\quad To isolate the RT instability in the $z$-direction, only $Z$ and $Q_r$ were perturbed, leading to the following dispersion relation:
\begin{equation}
\pi(2+\alpha^2)\omega^2+2\pi\alpha^2 = 0.
\label{nd_zRT}
\end{equation}
Eq.~(\ref{nd_zRT}) is quadratic with no real roots, confirming that this mode is stable and does not independently contribute to node formation. Consequently, we confirm that curvature effects play a negligible role in the rim instability mechanism, thereby justifying the use of the planar-crown assumption in our analysis.

\clearpage

\noindent{\bf Declaration of Interests:} The authors report no conflict of interest. \\
\\
\noindent{\bf Acknowledgements:} {S.B. acknowledges ANRF for the financial support through the grant ANRF/IRG/2024/000711/ENS and the computational resources provided by PARAM Seva under the National Supercomputing Mission, Government of India. K.C.S. thanks IIT Hyderabad for the financial support through the grant IITH/CHE/F011/SOCH1. We are deeply grateful to the anonymous reviewers for their valuable suggestions, which enhanced the quality and depth of the manuscript.}


\end{document}